\documentclass{emulateapj}
\usepackage{color}
\usepackage{courier}
\usepackage{sidecap}
\usepackage{longtable}
\usepackage{booktabs}
\usepackage{changepage}
\usepackage{amsmath}

\def\simlt{\lower.5ex\hbox{$\; \buildrel < \over \sim \;$}}
\def\simgt{\lower.5ex\hbox{$\; \buildrel > \over \sim \;$}}

\def\gsim{\lower 2pt \hbox{$\, \buildrel {\scriptstyle >}\over
{\scriptstyle \sim}\,$}}
\def\lsim{\lower 2pt \hbox{$\, \buildrel {\scriptstyle <}\over
{\scriptstyle \sim}\,$}}

\def\deg{\ifmmode ^{\circ}
         \else $^{\circ}$\fi}
\def\pdeg{\ifmmode
           $\setbox0=\hbox{$^{\circ}$}\rlap{\hskip.11\wd0 .}$^{\circ}
     \else \setbox0=\hbox{$^{\circ}$}\rlap{\hskip.11\wd0 .}$^{\circ}$\fi}
     
\def\pc{\ifmmode \mathrm{pc} \else $\mathrm{pc}$ \fi}
\def\mpc{\ifmmode \mathrm{Mpc} \else $\mathrm{Mpc}$\fi}
\def\mpcthree{\ifmmode \mathrm{Mpc}^{-3} \else $\mathrm{Mpc}^{-3}$\fi}
\def\gpcthree{\ifmmode \mathrm{Gpc}^{-3} \else $\mathrm{Gpc}^{-3}$\fi}

\def\kelvin{\ifmmode \mathrm{K} \else {$\mathrm{K}$}\fi}
\def\kev{\ifmmode \mathrm{keV} \else $\mathrm{keV}$ \fi}

\def\lsun{\ifmmode {L_\odot} \else $L_\odot$\fi}
\def\msun{\ifmmode M_\odot \else $M_\odot$\fi}
\def\msunyr{\ifmmode M_\odot~\mathrm{yr}^{-1} \else $M_\odot~\mathrm{yr}^{-1}$\fi}

\def\cosi{\ifmmode {\cos\,i} \else $\cos\,i$\fi}

\def\heii{\ifmmode {\rm He{\sc ii}} \else He~{\sc ii}\fi}
\def\mgii{\ifmmode {\rm Mg{\sc ii}} \else Mg~{\sc ii}\fi}
\def\caii{\ifmmode {\rm Ca{\sc ii}} \else Ca~{\sc ii}\fi}
\def\ciii{\ifmmode {\rm C{\sc iii}]} \else C~{\sc iii}]\fi}
\def\civ{\ifmmode {\rm C{\sc iv}} \else C~{\sc iv}\fi}
\def\mgii{\ifmmode {\rm Mg{\sc ii}} \else Mg~{\sc ii}\fi}

\def\teff{\ifmmode {T_{\rm eff}} \else $T_{\rm eff}$\fi}
\def\tmax{\ifmmode {T_{\rm max}} \else $T_{\rm max}$\fi}

\def\mbh{\ifmmode {M_{\rm BH}} \else $M_{\rm BH}$\fi}
\def\led{\ifmmode L_{\mathrm{Ed}} \else $L_{\mathrm{Ed}}$\fi}
\def\lbolflare{\ifmmode L_{\mathrm{bol,flare}} \else $L_{\mathrm{bol,flare}}$\fi}
\def\lagn{\ifmmode L_{\mathrm{agn}} \else $L_{\mathrm{agn}}$\fi}
\def\lbolagn{\ifmmode L_{\mathrm{bol,agn}} \else $L_{\mathrm{bol,agn}}$\fi}
\def\lbol{\ifmmode L_{\mathrm{bol}} \else $L_{\mathrm{bol}}$\fi}
\def\mdot{\ifmmode {\dot M} \else $\dot M$\fi}
\def\mdoto{\ifmmode {\dot{M}_0} \else  $\dot{M}_0$\fi}
\def\mdotf{\ifmmode {\dot{M}_\mathrm{flare}} \else  $\dot{M}_\mathrm{flare}$\fi}

\def\hnot{\ifmmode H_0 \else H$_0$ \fi}

\def\vkep{\ifmmode v_\mathrm{Kep} \else $v_\mathrm{Kep}$ \fi}
\def\vc{\ifmmode v_\mathrm{c} \else $v_\mathrm{c}$ \fi}

\def\vthree{\ifmmode v_{1000} \else $v_{1000}$ \fi}
\def\vrel{\ifmmode v_\mathrm{rel} \else $v_\mathrm{rel}$ \fi}
\def\vkick{\ifmmode v_\mathrm{kick} \else $v_\mathrm{kick}$ \fi}
\def\vkickz{\ifmmode v_{\mathrm{kick},z} \else $v_{\mathrm{kick},z} $ \fi}
\def\vkicky{\ifmmode v_{\mathrm{kick},y} \else $v_{\mathrm{kick},y} $ \fi}
\def\vchar{\ifmmode v_\mathrm{char} \else $v_\mathrm{char}$ \fi}
\def\eflare{\ifmmode E_\mathrm{flare} \else $E_\mathrm{flare}$ \fi}
\def\ekick{\ifmmode E_\mathrm{kick} \else $E_\mathrm{kick}$ \fi}
\def\ecoll{\ifmmode E_\mathrm{coll} \else $E_\mathrm{coll}$ \fi}
\def\ezero{\ifmmode E_\mathrm{0} \else $E_\mathrm{0}$ \fi}
\def\efac{\ifmmode \xi_\mathrm{E} \else $\xi_\mathrm{E}$ \fi}
\def\tqso{\ifmmode t_\mathrm{QSO} \else $t_\mathrm{QSO}$ \fi}
\def\tflare{\ifmmode t_\mathrm{flare} \else $t_\mathrm{flare}$ \fi}
\def\tzero{\ifmmode t_\mathrm{0} \else $t_\mathrm{0}$ \fi}
\def\tfac{\ifmmode \xi_\mathrm{t} \else $\xi_\mathrm{t}$ \fi}
\def\gfac{\ifmmode f_\mathrm{g} \else $f_\mathrm{g}$ \fi}
\def\lflare{\ifmmode L_\mathrm{flare} \else $L_\mathrm{flare}$ \fi}
\def\fflare{\ifmmode F_\mathrm{flare} \else $F_\mathrm{flare}$ \fi}
\def\nflare{\ifmmode N_\mathrm{flare} \else $N_\mathrm{flare}$ \fi}
\def\tshock{\ifmmode T_\mathrm{shock} \else $T_\mathrm{shock}$ \fi}
\def\rmin{\ifmmode R_\mathrm{1} \else $R_\mathrm{1}$ \fi}
\def\rmax{\ifmmode R_\mathrm{2} \else $R_\mathrm{2}$ \fi}
\def\rbound{\ifmmode R_\mathrm{b} \else $R_\mathrm{b}$ \fi}
\def\pbound{\ifmmode P_\mathrm{b} \else $P_\mathrm{b}$ \fi}
\def\mbound{\ifmmode M_\mathrm{b} \else $M_\mathrm{b}$ \fi}
\def\mbo{\ifmmode M_{\mathrm{b}0} \else $M_{\mathrm{b}0} $ \fi}
\def\ebo{\ifmmode E_{\mathrm{b}0} \else $E_{\mathrm{b}0} $ \fi}
\def\efinal{\ifmmode E_\mathrm{final} \else $E_\mathrm{final} $ \fi}
\def\tbound{\ifmmode t_\mathrm{b} \else $t_\mathrm{b}$ \fi}
\def\tagn{\ifmmode t_\mathrm{AGN} \else $t_\mathrm{AGN}$ \fi}
\def\torb{\ifmmode t_\mathrm{orb} \else $t_\mathrm{orb}$ \fi}
\def\tdf{\ifmmode t_\mathrm{df} \else $t_\mathrm{df}$ \fi}
\def\rlim{\ifmmode R_\mathrm{lim} \else $R_\mathrm{lim}$ \fi}
\def\vlim{\ifmmode v_\mathrm{lim} \else $v_\mathrm{lim}$ \fi}
\def\vphi{\ifmmode v_\phi \else $v_\phi$ \fi}
\def\mlim{\ifmmode M_\mathrm{lim} \else $M_\mathrm{lim}$ \fi}
\def\tlim{\ifmmode t_\mathrm{lim} \else $t_\mathrm{lim}$ \fi}
\def\llim{\ifmmode L_\mathrm{lim} \else $L_\mathrm{lim}$ \fi}
\def\fqso{\ifmmode f_\mathrm{QSO} \else $f_\mathrm{QSO}$ \fi}

\def\hbeta{\ifmmode \rm{H}\beta \else H$\beta$\fi}
\def\hbetan{\ifmmode \rm{H}\beta_{\rm n} \else H$\beta_{\rm n}$\fi}
\def\hgamma{\ifmmode \rm{H}\gamma \else H$\gamma$\fi}
\def\hdelta{\ifmmode \rm{H}\delta \else H$\delta$\fi}
\def\hepsilon{\ifmmode \rm{H}\epsilon \else H$\epsilon$\fi}
\def\hzeta{\ifmmode \rm{H}\zeta \else H$\zeta$\fi}
\def\halpha{\ifmmode \rm{H}\alpha \else H$\alpha$\fi}
\def\lalpha{\ifmmode \rm{Ly}\alpha \else Ly$\alpha$}

\def\dvhb{\ifmmode \Delta v_{\hbeta} \else $\Delta v_{\hbeta}$\fi}
\def\dvmg{\ifmmode \Delta v_{\rm{Mg}} \else $\Delta v_{\rm{Mg}}$\fi}

\def\muobs{\ifmmode {\mu_{o}} \else  $\mu_{o}$ \fi}
\def\cosi{\ifmmode {\mathrm{cos}\,i} \else $\mathrm{cos}\,i$\fi}

\def\teff{\ifmmode {T_{eff}} \else $T_{eff}$ \fi}
\def\tmax{\ifmmode {T_{max}} \else $T_{max}$ \fi}

\def\tauh{\ifmmode {\tau_{\rm H}} \else $\tau_{\rm H}$ \fi}

\def\yr{\ifmmode {\rm yr} \else  yr \fi}
\def\kms{\ifmmode \rm km~s^{-1}\else $\rm km~s^{-1}$\fi}
\def\cm{\ifmmode {\rm cm} \else  cm \fi}
\def\cmmitwo{\ifmmode \rm cm^{-2} \else $\rm cm^{-2}$\fi}
\def\cmmithree{\ifmmode \rm cm^{-3} \else $\rm cm^{-3}$\fi}
\def\cmps{\ifmmode \rm cm~s^{-1}\else $\rm cm~s^{-1}$\fi}
\def\cmpsps{\ifmmode \rm cm~s^{-2}\else $\rm cm~s^{-2}$\fi}
\def\kmps{\ifmmode \rm km~s^{-1}\else $\rm km~s^{-1}$\fi}
\def\kmpspmpc{\ifmmode \rm km~s^{-1}~Mpc^{-1} \else
    $\rm km~s^{-1}~Mpc^{-1}$\fi}
  
\def\gcmthree{\ifmmode \rm g~cm^{-3} \else $\rm g~cm^{-3}$\fi}
\def\gcmtwo{\ifmmode \rm g~cm^{-2} \else $\rm g~cm^{-2}$\fi}
   
\def\erg{\ifmmode {\rm erg} \else $\rm erg$ \fi}
\def\ergps{\ifmmode {\rm erg~s^{-1}} \else $\rm erg~s^{-1}$ \fi}
\def\ergcms{\ifmmode \rm erg~cm^{-2}~s^{-1} \else $\rm erg~cm^{-2}~s^{-1}$ \fi}
\def\ergcmshz{\ifmmode \rm erg~s^{-1}~cm^{-2}~Hz^{-1} \else $\rm
erg~cm^{-2}~s^{-1}~Hz^{-1}$ \fi}
\def\ergcmsa{\ifmmode \rm erg~cm^{-2}~s^{-1}~\AA^{-1} \else $\rm
erg~cm^{-2}~s^{-1}~\AA^{-1}$ \fi}
\def\ergshz{\ifmmode \rm erg s^{-1} Hz^{-1} \else
   $\rm erg s^{-1} Hz^{-1}$ \fi}

\def\lam{\ifmmode {\lambda} \else {$\lambda$} \fi}
\def\llam{\ifmmode {L_\lambda} \else  $L_\lambda$ \fi}
\def\lamLlam{\ifmmode \lambda L_{\lambda}(5100) \else {$\lambda L_{\lambda}(5100)$} \fi}
\def\nuLnu{\ifmmode \nu L_{\nu}(5100) \else {$\nu L_{\nu}(5100)$} \fi}
\def\ilam{\ifmmode {I_\lambda} \else  $I_\lambda$ \fi}
\def\flam{\ifmmode {F_\lambda} \else  $F_\lambda$ \fi}
\def\inu{\ifmmode {I_\nu} \else  $I_\nu$ \fi}
\def\fnu{\ifmmode {F_\nu} \else  $F_\nu$ \fi}
\def\bnu{\ifmmode {B_\nu} \else  $B_\nu$ \fi}

\def\msigma{\ifmmode M_{\sigma} \else $M_{\sigma}$\fi}
\def\mbulge{\ifmmode M_{\mathrm{bulge}} \else $M_{\mathrm{bulge}}$\fi}
\def\mgal{\ifmmode M_{\mathrm{gal}} \else $M_{\mathrm{gal}}$\fi}
\def\lgal{\ifmmode L_{\mathrm{gal}} \else $L_{\mathrm{gal}}$\fi}
\def\lbulge{\ifmmode L_{\mathrm{bulge}} \else $L_{\mathrm{bulge}}$\fi}
\def\mgalstar{\ifmmode M^*_{\mathrm{gal}} \else $M^*_{\mathrm{gal}}$\fi}

\def\mbhsigstar{\ifmmode M_{\mathrm{BH}} - \sigma_* \else $M_{\mathrm{BH}} - \sigma_*$ \fi}
\def\deltalogmbh{\ifmmode \Delta~{\mathrm{log}}~M_{\mathrm{BH}} \else $\Delta$~log~$M_{\mathrm{BH}}$\fi}

\def\sigstar{\ifmmode \sigma_* \else $\sigma_*$\fi}
\def\sigthree{\ifmmode \sigma_{\mathrm{[O~III]}} \else $\sigma_{\mathrm{[O~III]}}$\fi}
\def\sigtwo{\ifmmode \sigma_{\mathrm{[O~II]}} \else $\sigma_{\mathrm{[O~II]}}$\fi}
\def\signl{\ifmmode \sigma_{\mathrm{NL}} \else $\sigma_{\mathrm{NL}}$\fi}
\def\wthree{\ifmmode {\rm FWHM({[O~III]})} \else $FWHM({[O~III]})$ \fi}
\def\wtwo{\ifmmode {\rm FWHM({[O~II]})} \else $FWHM({[O~II]})$ \fi}
\def\mthree{\ifmmode M_{\mathrm [O~III]} \else $M_{\mathrm [O~III]}$ \fi}
\def\mtwo{\ifmmode M_{\mathrm [O II]} \else $M_{\mathrm [O II]}$ \fi}
\def\lbreak{\ifmmode L_{\mathrm{break}} \else $L_{\mathrm{break}}$\fi}
\def\lcut{\ifmmode L_{\mathrm{cut}} \else $L_{\mathrm{cut}}$\fi}

\shortauthors{Smith et al.}
\shorttitle{\emph{Swift} Survey of the \emph{Kepler} Field}

\begin{document}

\title{KSwAGS: A \emph{Swift} X-ray and UV Survey of the \emph{Kepler} Field. I.}

\author{Krista Lynne Smith\altaffilmark{1,2}, Patricia~T.~Boyd\altaffilmark{2}, Richard~F.~Mushotzky\altaffilmark{1}, Neil~Gehrels\altaffilmark{2}, Rick~Edelson\altaffilmark{1}, Steve~B.~Howell\altaffilmark{3}, Dawn~M.~Gelino\altaffilmark{4}, Alexander~Brown\altaffilmark{5} \& Steve~Young\altaffilmark{1}}

\altaffiltext{1}{Department of Astronomy, University of Maryland College Park; klsmith@astro.umd.edu}

\altaffiltext{2}{NASA/GSFC, Greenbelt, MD 20771, USA}

\altaffiltext{3}{NASA Ames Research Center, Moffett Field, CA 94095, USA}

\altaffiltext{4}{NASA Exoplanet Science Institute, Caltech, Pasadena, CA 91125, USA}

\altaffiltext{5}{CASA, University of Colorado, Boulder, CO 80309, USA}

\begin{abstract}

We introduce the first phase of the \emph{Kepler}-\emph{Swift} Active Galaxies and Stars survey (KSwAGS), a simultaneous X-ray and UV survey of $\sim$6 square degrees of the \emph{Kepler} field using the \emph{Swift} XRT and UVOT. We detect 93 unique X-ray sources with S/N $\geq$3 with the XRT, of which 60 have UV counterparts. We use the Kepler Input Catalog (KIC) to obtain the optical counterparts of these sources, and construct the $f_X / f_V$ ratio as a first approximation of the classification of the source. The survey produces a mixture of stellar sources, extragalactic sources, and sources which we are not able to classify with certainty. We have obtained optical spectra for thirty of these targets, and are conducting an ongoing observing campaign to fully identify the sample. For sources classified as stellar or AGN with certainty, we construct SEDs using the 2MASS, UBV and GALEX data supplied for their optical counterparts by the KIC, and show that the SEDs differ qualitatively between the source types, and so can offer a method of classification in absence of a spectrum. Future papers in this series will analyze the timing properties of the stars and AGN in our sample separately. Our survey provides the first X-ray and UV data for a number of known variable stellar sources, as well as a large number of new X-ray detections in this well-studied portion of the sky. The KSwAGS survey is currently ongoing in the K2 ecliptic plane fields. 

\end{abstract}

\section{Introduction}
\label{sec:intro}

The \emph{Kepler} mission was designed to detect exoplanets in the habitable zone by searching for repeating transits in the light curves of over 150,000 sunlike stars. The exceptional photometric and temporal precision, high duty cycle and rapid, continuous sampling required for this task make \emph{Kepler} a unique asset for the study of various other astrophysical targets in its 100 square-degree field of view (FOV). To detect exoplanets with up to year-long orbital timescales, \emph{Kepler} remained continuously pointed at a region of the sky in the constellation Cygnus, chosen for its high density of observable dwarf stars. During its prime mission, \emph{Kepler} collected hundreds of thousands of light curves for sources over baselines of a few months to 4.25 years by telemetering ``postage stamps" of pixels around chosen stars and sampling with 30-minute cadence. In addition to high precision light curves, \emph{Kepler} also produces Full Frame Images (FFIs), 29.4-minute exposures of the entire FOV, approximately once per month. The \emph{Kepler} Science Center maintains that the full field contains $\sim$10 million stars above the confusion limit of 20-21 mag, and the FFIs can provide photometry on monthly cadences for all objects in the \emph{Kepler} FOV, regardless of whether high-cadence light curves were collected. This rich data set includes previously unidentified variable astrophysical sources  over an impressive range of distance scales, both within the Galaxy and throughout the larger universe. Sources within the Milky Way include RR Lyrae stars \citep{gugg12}, rapidly oscillating peculiar A (roAp) stars \citep{balona13}, and cataclysmic variables (CVs) \citep{scaringi13}. 

Optically variable extragalactic sources such as Seyfert~1 galaxies \citep{mush11} and BL~Lac objects \citep{edelson13} are tantalizing candidates, since the accretion physics in active galactic nuclei (AGN) is poorly understood. The size scale of the accretion disk is approximately $\leq0.01$~parsecs, or $\sim$1~milliarcsec for even the nearest active galaxies. Such a measurement is still well below resolvable scales for any optical observatories. Variability studies are then the only direct probe of conditions within the disk itself, and consequently of the process of accretion. Theory predicts model-dependent light curve characteristics;  for example, \citet{reynolds09} predicted clear g-mode oscillations and characteristic frequencies in the power spectrum that correspond to local acoustic waves. One also expects that the characteristic timescales of the disk (i.e., the thermal, viscous, and dynamical timescales) would be evident in light curve with sufficiently rapid and regular sampling. \emph{Kepler} provides the AGN community with its first opportunity to measure such effects in the optical. 

\emph{Kepler}'s bounty of unprecedented high-precision light curves has resulted in its FOV being one of the best-studied regions of the sky. Objects in the KIC (\emph{Kepler} Input Catalog) overlap with the 2-Micron All-Sky Survey  \citep[2MASS;][]{skrutskie06}, the Wide-field Infrared Survey Explorer \citep[WISE;][]{wright10}, and the Galaxy Evolution Explorer survey  \citep[GALEX;][]{martin03}. Additionally, \citet{everett12} conducted an optical photometric survey of the field in the $UBV$ bands. An X-ray survey of the Kepler field is therefore prudent, providing an important and unique resource for locating interesting variable objects in this field.

X-ray selection is an effective way to curate a sample of astrophysically interesting variable stars and extragalactic sources. To capitalize on the wealth of data for the objects in the \emph{Kepler} field, both photometric and archival, we have conducted the \emph{Kepler-Swift} Active Galaxies and Stars survey (KSwAGS), a \emph{Swift} X-ray telescope (XRT) survey of four modules of the \emph{Kepler} field that is approximately ten times deeper than the ROSAT All-Sky Survey \citep[RASS;][]{voges99}. We chose these four modules to lie approximately perpendicular to the galactic plane, sampling a range of galactic latitudes. Additionally, the co-aligned \emph{Swift} UV/Optical Telescope (UVOT) provides concurrent UV coverage for each KSwAGS pointing. We present here the first KSwAGS catalog of X-ray sources in the original \emph{Kepler} FOV, their UVOT and archival data, and an introduction to the KSwAGS series of temporal analysis papers. This first phase of the KSwAGS survey detected 93 unique X-ray sources. This includes a number of known stellar variables and two previously identified AGN. For newly discovered sources, we present the best determination of the nature of the object via optical spectra, X-ray to optical flux ratios ($f_X / f_V$), the shapes of the broadband spectral energy distribution (SEDs), and, where possible, inspection of the optical light curves. Since the \emph{Kepler} FOV was chosen explicitly to be typical of the Milky Way galaxy, results should be widely relevant and offer a typical density of exotic variable sources near the galactic plane. This paper serves as an introduction to the survey, which is currently ongoing in the new \emph{Kepler} (known as K2) ecliptic plane fields. The time variability physics and temporal analyses will be presented in two upcoming papers, focused separately on the stellar sources and the AGN.

This paper is organized as follows: in Section~\ref{sec:selection}, we describe our X-ray and UV survey and the identification of the optical counterparts. Section~\ref{sec:class} explains the various methods used to classify the sources as either stellar or extragalactic, and categorize the sources based on their optical spectra or archival information. In Section~\ref{sec:lcs}, we provide samples of the light curves and analysis of a stellar source and an AGN, to exemplify the content of the follow-up papers dedicated to each of these samples. Section~\ref{sec:k2} describes our ongoing survey in the new ecliptic plane K2 fields, and the final section summarizes the products of the survey and future directions. 

\section{Survey Field and Source Detection}
\label{sec:selection}
 
KSwAGS was conducted with the \emph{Swift} X-ray Telescope (XRT), which operates in the 0.2~-~10~keV range with a sensitivity of $2\times10^{-14}$ erg cm$^{-2}$ s$^{-1}$ in $10^{4}$ seconds. The XRT field of view (FOV) is 23.6$\times$23.6~arcmin \citep{burrows05}. The co-aligned UV/Optical Telescope (UVOT) is sensitive to the 170-650 nm range and has a sensitivity of $B=22.3$ in white light in 1000 seconds and a FOV of 17$\times$17 arcmin \citep{roming05}. The UVOT has six broad-band optical and UV filters; the KSwAGS survey uses the the near-UV  \emph{uvm2} filter ($\lambda_{c}=2246$~\AA), since the \emph{Kepler} FOV is already well-studied in the optical. Additionally, the UVOT NUV filters suffer from a phenomenon known as ``red leak," in which the NUV flux is artificially inflated by the counting of optical photons due to red tails in the filters' transmission curves. The \emph{uvm2} filter has the smallest amount of  red leak of the three NUV filters \citep{breeveld11}. 

\begin{figure}[ht]
\begin{center}
\plotone{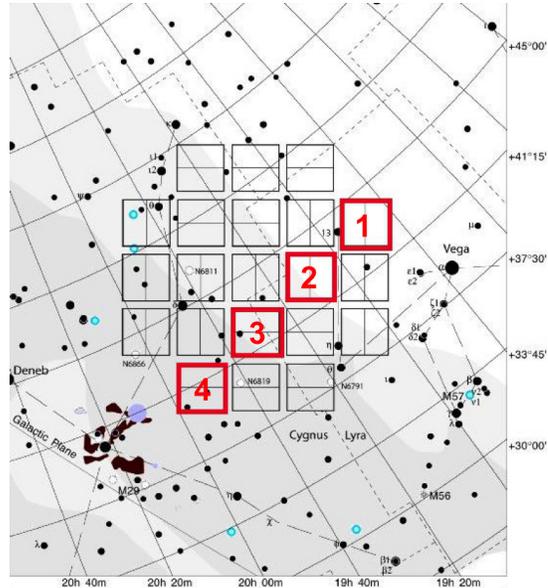}
\figcaption[]{Map of the \emph{Kepler} FOV, with the \emph{Swift}-surveyed modules outlined in red. Region 1 is the upper right module, with the regions increasing in number toward the galactic plane (bottom left).
\label{fig:modules}}
\end{center}
\end{figure}

\begin{figure*}[ht]
\begin{center}
\includegraphics[width=0.7\textwidth,height=7cm]{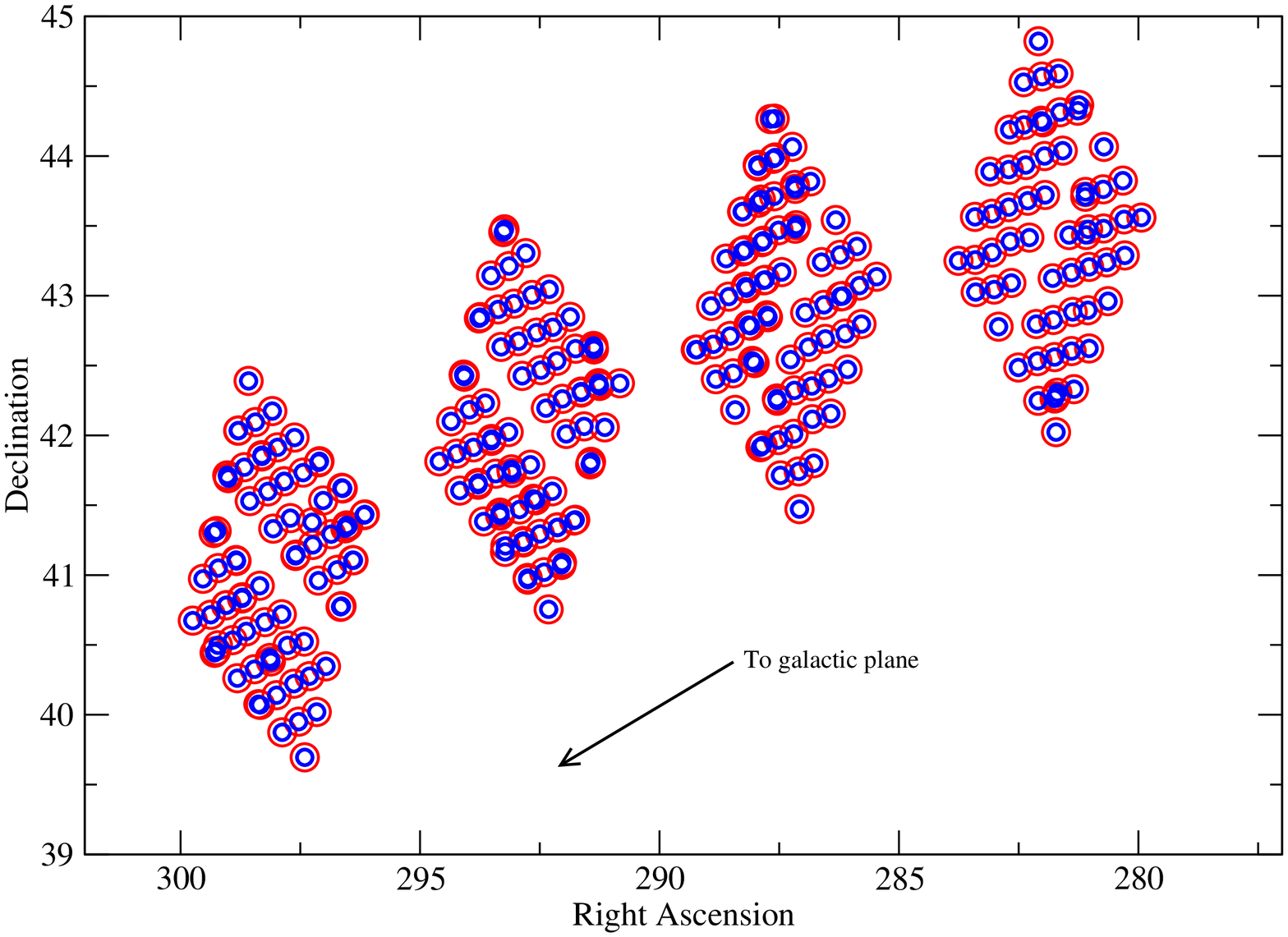}
\figcaption{Arrangement of our XRT pointings (red) and UVOT pointings (blue). The UVOT FOV is approximately half the size of the XRT FOV. Overlapping circles occur when multiple pointings were required to obtain the full 2ks observation. The right ascension axis is inverted for easier comparison to Figure~\ref{fig:modules}. Note that the shape of the UVOT FOV is actually roughly square, but is oriented at various angles within the larger XRT FOV. We use circles here to illustrate the approximate difference in UV/X-ray coverage.
\label{fig:mosaic}}
\end{center}
\end{figure*}

The entire \emph{Kepler} FOV consists of 21 modules. KSwAGS covers four of these modules, outlined in red in Figure~\ref{fig:modules}, and subtends $\sim$6 square degrees. The modules are roughly perpendicular to the galactic plane, decreasing in galactic latitude from Region~1 to Region~4. Each of the four modules is covered in 56 total pointings, indexed from 00 to 55, with each pointing lasting approximately 2 kiloseconds. Figure~\ref{fig:mosaic} shows the locations of the XRT and UVOT pointings on the sky; note the smaller UVOT field of view. 

For source detection, the XRT raw images and exposure maps were analyzed using \texttt{XIMAGE}. The background was optimized and the sources were located using the task \texttt{detect} with a S/N~$\geq3$, making use of the \texttt{bright} qualifier. The \texttt{bright} flag creates a weighted mean of each excess and optimizes the point-spread function (PSF), preventing the detection algorithm from erroneously interpreting one large source as several smaller ``detections"; however, note that if two \emph{real} sources were within the optimized PSF, our method would report only one source. 

After removing duplicate sources from pointing overlaps and eliminating detections of extended emission from single sources, we have detected 93 total X-ray sources with S/N~$\geq3$, which corresponds to approximately 12 total counts or 0.006~cts~s$^{-1}$ within our narrow range of exposure times and background estimates. Table \ref{tab:coords} lists these detections, as well each source's XRT count rate and exposure time. In this table, we assign each source an identifying number (KSw \#) based on increasing right ascension. We will use this ID number to refer to the sources throughout the paper and in subsequent tables. 

On the \emph{Swift} spacecraft, the XRT and UVOT are co-aligned to allow for simultaneous multiwavelength follow-up of GRBs detected onboard. To identify the counterpart in the simultaneous UVOT images, we overplotted a contour map of each XRT source onto the co-aligned UVOT image, and selected the counterpart visually. Once a counterpart is selected, we use the FTOOL \texttt{uvotsource} to obtain the UV source's position and flux density. This information is also given in Table \ref{tab:coords}. The UVOT FOV is approximately half the size of the XRT FOV, so some of our sources do not have UVOT counterparts listed because the XRT source was beyond the edge of the UVOT image. Note, however, that the overlap of our \emph{Swift} pointings (see Figure \ref{fig:mosaic}) mitigates this problem to some degree, and so $\sim65$\% of our XRT sources have measurable UVOT counterparts. 

Once all detectable UVOT counterparts had been identified, we looked up their optical counterparts in the \emph{Kepler} Input Catalog (KIC). For our typical S/N ratio, the XRT can typically localize a source position to 3.5$''$ with 90\% confidence. Much of this uncertainty is due to the star-tracker attitude solution onboard the spacecraft \citep{goad07}. Positional uncertainty can be reduced when a UVOT sky image is present. With a PSF of 2.5$''$ for each filter and a pixel scale of $\sim0.5''$, an uncorrected UVOT position is typically accurate to within 1 arcsecond \citep{roming05}. When stars in the UVOT field of view can be matched to the USNOB1 catalog, the aspect correction can result in UVOT positions that are accurate to within 0.5$''$.   Therefore, if the object has a UVOT counterpart, we use this position to query the KIC. An X-ray source might not be detected as UVOT source if it suffers from heavy extinction either from dust in the Milky Way or, in the case of AGN, from innate host galaxy dust. This is certainly the case for KSw~93, Cygnus~A, which is a very bright X-ray source but is also a Type 2 AGN, in which UV light from the bright nucleus cannot escape the dust in the circumnuclear torus. However, the majority of our sources without UVOT counterparts result from the relative FOV sizes between the XRT and the UVOT. Table~\ref{t:tab2} denotes which of these was the case. In the absence of a UVOT source, we query using the X-ray coordinates. The set of coordinates that was used to query the KIC is given for each source in Table~\ref{t:tab2}. In 16 cases, an object had more than one possible KIC counterpart within 5$''$; the KSwAGS source number of each of these targets is indicated by an asterisk in Table~\ref{t:tab2}. We have done a case-by-case analysis for each of these to determine which counterpart is the most likely. In the cases of KSw~2, 20, 35, 39, 55, 82, and 93, we have spectroscopically confirmed the listed KIC counterpart as an AGN (see Section~\ref{sec:optspec}). Since AGN are ubiquitously X-ray sources, we can be confident that this is the correct identification. In the case of KSw~85, the listed KIC counterpart is a known ROSAT source and active star, and KSw~35 is a spectroscopically-confirmed white dwarf. The remaining cases (KSw~19, 26, 37, 41, 45, 50, 72, and 77) are less certain; we determined the most likely KIC counterpart for these objects by overlaying UVOT and XRT contours on STSci-DSS~III images and choosing the optical source most congruent with the available contour sets. 
Once determined, KIC counterpart can then be used to obtain the flux ratio $f_X / f_V$ (Section \ref{sec:optratio}), magnitudes from various sky surveys in other wavebands to construct broadband spectral energy distributions (SEDs) (Section \ref{sec:sed}), and most importantly, all quarters of \emph{Kepler} time series data for the source (Section \ref{sec:lcs}).

Of all 93 sources, 23 were already identified in the literature. We obtained optical spectra for an additional 30 sources on the 200-inch Hale telescope at Palomar Observatory during August~26-28~2014. These spectra were obtained using the double beam spectrograph (DBSP) with a slit width of 1$"$, equipped with the D-55 dichroic filter to split light between the blue and red arms. The blue arm used a 1200 mm$^{-1}$ grating with R$\sim$7700 and covered 1500~\AA. The red arm used a 1200 mm$^{-1}$ grating with R$\sim$10,000 and covered 670~\AA. We observed at least two spectrophotometric stars each night and arc lamp exposures were obtained before each source exposure at the source location. Red spectra were wavelength calibrated with a HeNeAr lamp and blue spectra with a FeAr lamp. While one night was clear and provided stable seeing at $\sim1"$, our seeing deteriorated on the second night, restricting our observations to targets with optical magnitudes $M_V\leq17$. Data reduction was done using IRAF two- and one-dimensional routines for spectroscopic data and produced a final one-dimensional spectrum for each observation. Optical spectra are further discussed in Section~\ref{sec:optspec}.

\section{Characteristics of Sources}
\label{sec:class}

All sources are summarized in Table~\ref{t:tab2}, which contains their most likely KIC counterparts; all available fluxes from GALEX, the \citet{everett12} \emph{UBV} survey, and 2MASS; the X-ray to optical flux ratios; and whether or not the object has an archived \emph{Kepler} light curve. We also provide any previous identifications of the sources in the literature or from the online databases of SIMBAD and NED. KSwAGS has provided new X-ray and UV data for a highly diverse mix of stellar sources, including rapid rotators, pulsating variables, eclipsing binaries, and $\gamma$~Doradus, $\delta$~Scuti, and $\delta$~Cepheid type stars. We have also detected the known Seyfert~1~AGN~Zw~229-15, a catalogued BL Lacertae object, and the radio galaxy Cygnus~A. From the literature identifications and our optical spectra, we know with certainty the optical counterparts of 53 sources total (57\%). For 9 additional sources, we can confidently assess whether the counterpart is stellar or extragalactic using a combination of apparent magnitude (e.g., an object with $V\sim8$ is too bright to be an AGN), $f_X / f_V$, $U-B$ color (e.g., a $V\sim20$ source with a very high $f_X / f_V$ is most likely to be an AGN), or broadband SED shape.

\begin{figure}[ht]
\begin{center}
\includegraphics[width=0.53\textwidth]{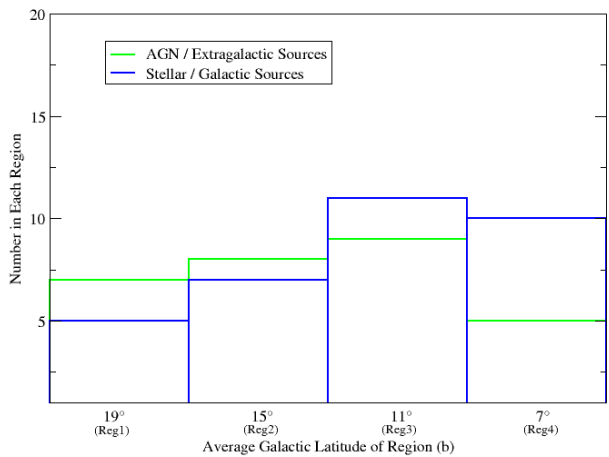}
\figcaption{Distribution of source types by average galactic latitude of each of our 4 regions. As shown in Figure~\ref{fig:modules}. Region numbers increase as they approach the galactic plane; Region 1 is furthest from the galactic plane with $b\sim19^{\circ}$, while Region 4 is the closest with $b\sim7^{\circ}$. Stellar sources begin to outnumber extragalactic sources closer to the plane, as expected. The plot includes the 62 sources classified by spectra, light curve behavior, or a combination of $f_X / f_V$ ratio, $U-B$ color, and apparent magnitude. Sources with unknown or uncertain classifications are not included in the plot.
\label{fig:fancy}}
\end{center}
\end{figure}

We expected that the relative fraction of stellar and extragalactic sources in a survey region would depend on the galactic latitude of the region. Figure~\ref{fig:fancy} shows the distribution in our four observing modules of the 62 sources for which we can confidently state a stellar or extragalactic origin. As the modules approach the galactic plane, the number of stellar sources outpaces the number of extragalactic sources, as expected. The diverse methods of classifying these sources prevents a robust analysis of the error; we display the trend only to illustrate the survey contents.

Below we outline the various methods used to classify the survey sources as either stellar or as AGN. Briefly: optical spectra are the most certain form of classification, but we do not have spectra for all KSwAGS targets. The X-ray to optical flux ratio offers a fairly stringent characterization at its extreme ends, but is degenerate at intermediate values. Broadband spectral energy distributions (SEDs) can be constructed for sources with adequate archival data in surveys at multiple wavelengths; the SED shape is recognizably different between stars and AGN and can be used to classify some sources. Finally, in the event that a source has very little archival data or an intermediate flux ratio value but does have a \emph{Kepler} light curve, one can use the temporal behavior to rule out an AGN in cases where the variability is strongly periodic. 

\subsection{Optical Spectra}
\label{sec:optspec}

\begin{figure*}
  \centering
  \setbox1=\hbox{\includegraphics[height=6in]{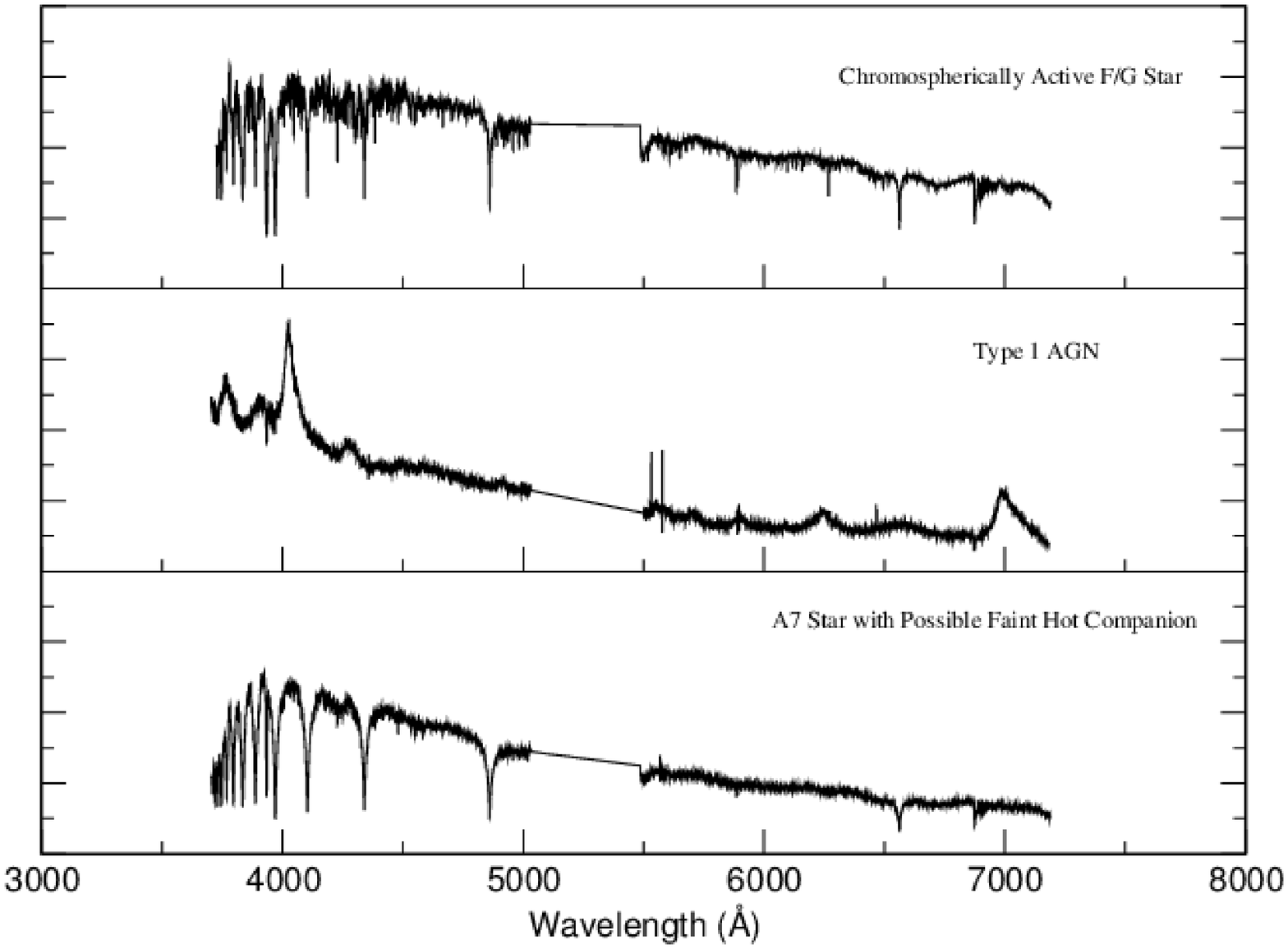}}
  \includegraphics[width=\textwidth,height=6in]{f4_back.eps}\llap{\makebox[15.6cm][l]{\raisebox{10.7cm}{\includegraphics[height=2cm]{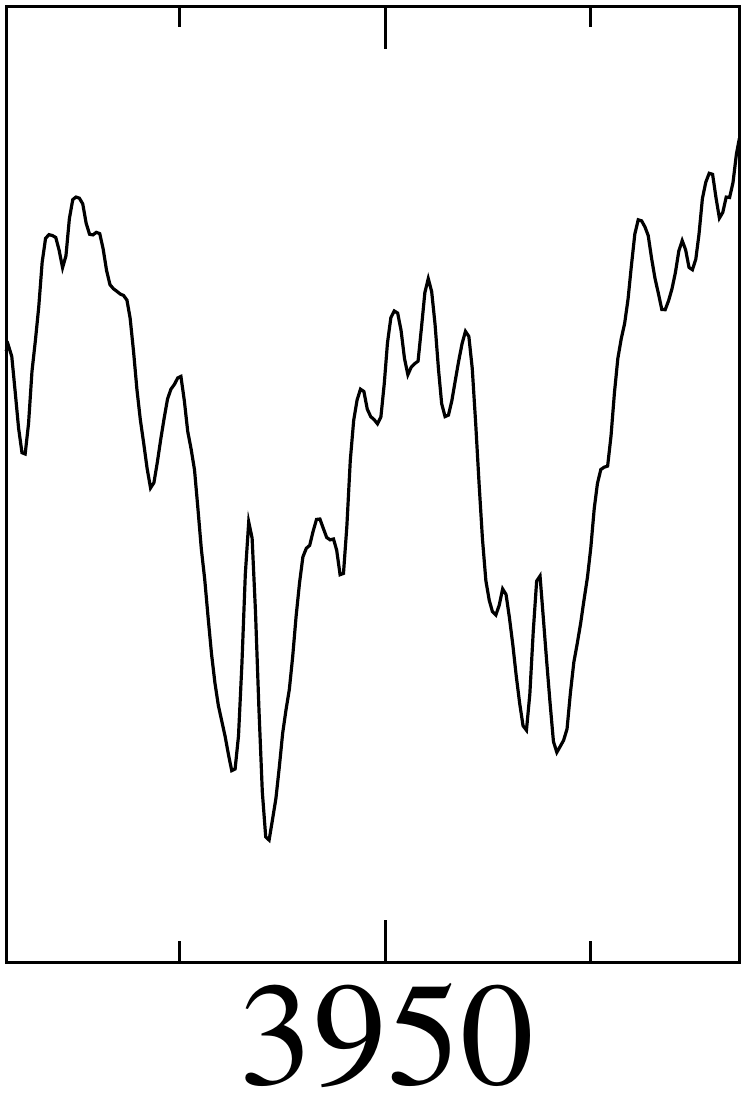}}}}
  \vspace{-30pt}
  \figcaption{Optical spectra from the 200-inch Hale telescope at Palomar observatory, for three representative KSwAGS objects. In order from top panel to bottom, they are KSw~19, KSw~55, and KSw~69. The inset in the top panel is an expanded view of the Ca H and K lines ($\lambda3969$\AA~and $\lambda3933$\AA), exhibiting the bright emission cores typical of chromospherically active stars.
  \label{fig:palomar}}
  \vspace{30pt}
\end{figure*}

We obtained simultaneous blue and red optical spectra for 31 sources in our sample with the double beam spectrograph (DBSP) on the 200-inch Hale telescope at Palomar Observatory. As described above, poor seeing on one of our nights restricted our observations to targets with $M_V \leq 17$, reducing the number of faint targets and disproportionately removing AGN from the observable sample; thus, most of our spectroscopic targets are stellar in nature. In all, the Palomar spectra identified 21 stellar sources and 10 AGN. Figure~\ref{fig:palomar} shows typical example spectra of three sources: a chromospherically active star, which are quite common in our sample; a type 1 AGN; and a normal A-type star. It is probable that the A star has a white dwarf companion that is producing the observed X-ray emission; alternatively, there may be confusion with a background X-ray source. Below we discuss the spectral properties of the AGN and stellar sources.

\subsubsection{AGN Spectra and Measured Parameters}
\label{sec:agnspec}

There are ten AGN among our Palomar spectra. Nine of them were type 1 (e.g., exhibit broad emission lines with FWHM$_{H_\beta} \geq 1000$~km/s), while one target, KSw~82, is a likely BL Lac object, exhibiting the typical flat, featureless continuum in multiple deep ($\sim$3200s) exposures. Newly confirmed AGN include KSw 2, 3, 25, 27, 39, 40, 55, 68, 82 and 92. They join the three previously known AGN in our sample: the BL Lac object BZB~J1848+4245 (KSw~9), the type 1 AGN Zw~229-15 (KSw~20), and the radio galaxy Cygnus~A (KSw~93). Their spectral types and redshifts are given in Table~\ref{t:tab3}. Our spectral analysis indicates that the KSwAGS AGN encompass a wide range of black hole masses ($7.3\leq \rm{log}~M_{BH}\leq9.4$), redshifts ($0.03\leq z\leq1.5$), and Eddington ratios ($0.003\leq\lambda_{Edd}\leq0.45$).

It has long been theorized that the characteristic optical variability timescale in the accretion disk of an AGN should correlate with the mass of the supermassive black hole. This relationship has already been demonstrated in the X-ray, where a break frequency in the power spectrum has been detected. The timescale associated with the break is assumed to correspond to the physical size of the accretion disk \citep{uttley02} and has been found to correlate with the black hole mass \citep{mchardy04}. For further discussion, see Section~\ref{sec:agnsample}. Recently, our group published the discovery of the first optical break frequency discovered in an AGN, using the \emph{Kepler} light curve of Zw~229-15 \citep{edelson14}. 

With this in mind, we calculate the redshift and black hole masses of our spectroscopically-confirmed AGN and present the values in Table~\ref{t:tab3}. We calculated the black hole masses using the formulae from \citet{wang09}; in the 7 objects that are low-redshift enough for the spectrum to contain H$\beta$, we use the relation:

\begin{align}
&\rm{log}\Big(\frac{M_{BH}}{10^6~M_\odot}\Big) = (1.39\pm0.14)+0.5~log~\Big(\frac{L_{5100}}{10^{44}~erg~s^{-1}}\Big) \nonumber \\
&\rm ~~~~~+ (1.09\pm0.23)~log~\bigg[\frac{FWHM(H\beta)}{1000~km~s^{-1}}\bigg].
\end{align}
For our three AGN with $z>0.6$, the H$\beta$ line is outside the clean region of the optical observing window, so we use the lower-wavelength \mgii~$\lambda2799$~\AA~line for BH mass estimation with the following relation:

\begin{align}
&\rm{log}\Big(\frac{M_{BH}}{10^6~M_\odot}\Big) = (1.13\pm0.27)+0.5~log~\Big(\frac{L_{3000}}{10^{44}~erg~s^{-1}}\Big) \nonumber \\
& \rm ~~~~~+ (1.51\pm0.49)~log~\bigg[\frac{FWHM(Mg \sc II)}{1000~km~s^{-1}}\bigg],
\end{align}

where $L_{3000}$ and $L_{5100}$ are the continuum luminosities at 3000~\AA~and 5100~\AA, respectively. The Eddington ratio $\lambda_{Edd}=L_{Edd}/L_{bol}$ is a measure of the observed luminosity compared to the theoretical luminosity output of the maximal spherical accretion rate (known as the Eddington luminosity). With the masses in hand, we may calculate the Eddington luminosity for each of the sources via $L_{Edd} = 1.38\times10^{38}~(M_{BH}/M_{\odot})$~erg~s$^{-1}$. The bolometric luminosity is estimated from the $0.2-10$ keV X-ray luminosity using the bolometric correction factor of $\sim$15 given by \citet{vasudevan07}; while this correction is widely used, there is significant empirical scatter in this relation, so the Eddington ratios given in Table~\ref{t:tab3} should be assumed with caution until detailed SED modeling is performed. The AGN types, redshifts, black hole masses and Eddington ratios are given for our confirmed AGN in Table~\ref{t:tab3}. In an upcoming paper, we will determine whether these AGN properties correlate with variability characteristics such as amplitude and characteristic frequency. A sample of the ongoing temporal analysis is given in Section~\ref{sec:lcs}.

\subsubsection{Stellar Spectra}
\label{sec:starspec}

The Palomar spectra of the KSwAGS survey stars fall into roughly three groups. The first is comprised of typical M-dwarf spectra with strong Ca~HK emission lines. M dwarfs with strong chromospheric activity are known to be X-ray sources, with high activity levels generally attributed to rapid rotation. Such rotation can indicate that the M dwarf is young and retains some innate formation spin, or that the star is in a binary with spin-orbit coupling \citep{lepine11}. There is also a population of normal main-sequence spectra, which are not indicative of X-ray activity. In these cases, it is possible that there is a dim white dwarf companion contributing the X-ray flux that is too faint to affect the optical spectrum. Finally, there is a large population of G through K stars that exhibit broad and complex emission structures in their H alpha lines and the Ca~HK lines (see top panel of Figure~\ref{fig:palomar}). The effective surface temperatures and surface gravities from the KIC for the stellar sources are given in Table~\ref{t:tab4}; also given are the spectral types gathered from the Palomar spectra. The spectral type can be inferred from the effective temperature even for those objects without Palomar spectra; however, note that the KIC temperatures are derived from $griz$ photometry and may be $\sim100-200$~K too low in many cases \citep{pinson12}.

\subsection{X-ray to Optical Ratios}
\label{sec:optratio}
When an X-ray source has a UV/optical counterpart, the calculation of the X-ray to optical flux ratio can provide source classification information. As noted in \citet{maccacaro88} for the Einstein Extended Medium Sensitivity Survey  \citep[EMSS;][]{gioia90} sources, the X-ray to optical flux ratio is a powerful tool in optical identification of X-ray sources. In general, all classes of AGN, including BL Lac objects, Seyfert galaxies and QSOs, have by far the highest $f_X / f_V$ ratios, followed by stellar accreting binaries. Coronal stellar sources such as dMe stars typically have the lowest ratios. \citet{krautter99} measured $f_X / f_V$ average values across a range of objects for a representative region of the RASS. Their average ratio values are $log[f_X/f_V] = -2.46\pm1.27$ for stellar sources and $log[f_X/f_V] = +0.41\pm0.65$ for AGN. 

\begin{figure*}[ht]
\begin{center}
\includegraphics[width=0.8\textwidth]{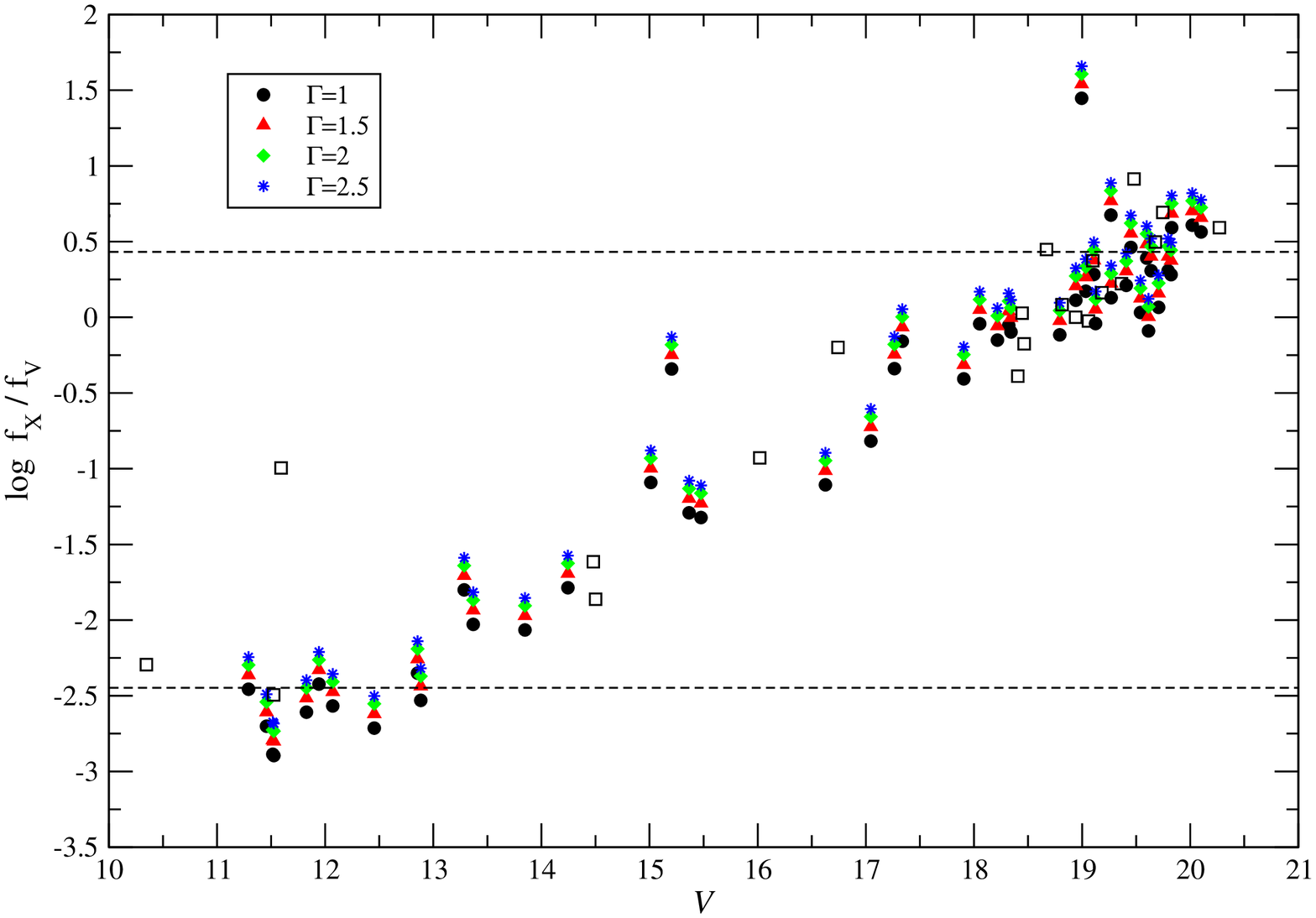}
\figcaption[]{X-ray and optical flux ratios for all sources in the sample with a \emph{V} magnitude given by the \citet{everett12} survey. Lines indicate the average value of this ratio for all stellar objects, $\langle log[f_X / f_V]\rangle = -2.46\pm1.27$, and for all AGN types, $\langle log[f_X / f_V]\rangle = 0.41\pm0.65$, as given in \citet{krautter99}. The X-ray fluxes were estimated using a variety of values for the photon index ($\Gamma$), shown in the legend. Black squares indicate objects for which no $V$ magnitude is available, and so the Kepler magnitude was used as a rough proxy. These values should be considered very approximate.
\label{fig:optrat}}
\end{center}
\end{figure*}

The optical counterparts for KSwAGS sources are drawn from the Kepler Input Catalog (KIC) (for details on counterpart identification, see Section \ref{sec:selection}). We draw our $V$ magnitudes used in constructing these ratios from the \citet{everett12} $UBV$ optical survey of the \emph{Kepler} FOV; the fluxes in these bands for objects covered by the $UBV$ survey are provided in Table~\ref{t:tab2}.

Typical X-ray detections have too few counts per energy bin to model the X-ray spectrum; therefore, in order to obtain the X-ray flux from the XRT count rates, one must assume a photon index that fits the general continuum shape of the underlying spectrum. Since we do not know the source type, we instead use PIMMS \citep{mukai93} to calculate the flux of each object using a range of feasible photon indices, from $\Gamma = 1$ to 2.5. This range encompasses typical values of $\Gamma$ for all object types, from active stellar sources to AGN. \citet{heinke08} found that magnetic and nonmagnetic CVs from the ASCA X-ray survey have average photon indices of $\Gamma=1.22\pm0.33$ and $1.97\pm0.20$, respectively. Values of $\Gamma$ for AGN have been measured for many samples and range from $\Gamma\sim1.5-2.5$ \citep[e.g.][]{nandra94,page05}. In all PIMMS count-to-flux conversions, we use the galactic column density at the source position cataloged by \citet{kalberla05}. We do not find that the choice of photon index changes the value of $f_X/f_V$ significantly. 

In Figure~\ref{fig:optrat}, we show the $f_X / f_V$ ratio of all survey sources that have an optical counterpart in the KIC within 5 arcseconds of the UVOT position (or the XRT position, if the object was outside the UVOT FOV or had no UVOT counterpart). For objects without $V$ magnitudes in the KIC, we use the \emph{Kepler} magnitude (given in Table~\ref{t:tab2} as ``kepmag") as a very rough proxy. The \emph{Kepler} magnitude is a generic optical magnitude calculated using various available optical measurements for any given object in the FOV; its detailed determination can be found in \citet{brown11}. Denoted in the figure are the average values for extragalactic and galactic sources from \citet{krautter99}. The KSwAGS survey conforms to these typical ratio values: the average value for all confirmed stellar objects in our survey for which optical data was available is $log[f_X/f_V] = -2.09\pm0.27$; for confirmed AGN, the value is $log[f_X/f_V] = +0.44\pm0.81$. The calculated values of the ratio for each object are given in Table~\ref{t:tab2}. Figure~\ref{fig:optrat} shows that the full distribution of KSwAGS sources cluster around the AGN and stellar average values, but have a significant spread. Indeed, as \citet{krautter99} points out, there is overlap between the two types of objects. In particular, white dwarfs and cataclysmic binaries show very high $f_X / f_V$ ratios for stellar sources, reaching into the lower tail of the AGN range. Therefore, in absence of a spectrum, a low or high $f_X / f_V$ value will at least offer a reliable classification of whether an X-ray source is stellar or extragalactic. Intermediate values of $f_X / f_V$ are not reliable classifiers, as there is significant overlap between source types for middling ratios.

\subsection{SEDs}
\label{sec:sed}

Using the multiple photometric data points provided from the 2MASS, GALEX, and $UBV$ \citet{everett12} surveys, along with our UVOT and XRT data, we can construct broadband spectral energy distributions (SEDs) for our sample across a wide wavelength range from infrared through X-ray. The archival surveys utilize varying magnitude systems, so in order to construct the SED, we converted these various systems to the consistent unit of millijanksies (mJy). The following discussion outlines this conversion process for each different survey. The final mJy fluxes are given in Table~\ref{t:tab2}.

\subsubsection{GALEX Magnitudes}
\label{sec:galex}

The GALEX satellite surveyed the sky in the FUV and NUV bands, with effective wavelengths of 1538.6 \AA~and 2315.7 \AA, respectively. GALEX data releases GR6 and GR7 included increased coverage of the \emph{Kepler} FOV, resulting in the KIC-GALEX crossmatched survey. There are two possible search mechanisms for the crossmatched survey. As defined by GALEX, they are 1) the ``accurate" search method, which only returns a counterpart that is within 2.5\arcsec~of the query coordinates and is unique (i.e., there are no other matches to either the KIC or GALEX source within this separation), and 2) the ``complete" search method, which returns all counterparts within 5\arcsec~of the query coordinates. In order to reduce spurious information in our survey, we have provided GALEX magnitudes only when our KIC coordinates have a GALEX counterpart using the ``accurate" method.

 The survey uses the AB magnitude system \citep{okegunn83}. The zero points of the GALEX magnitudes for the two bands were calculated by \citet{morrissey07} to be 18.82 (FUV) and 20.08 (NUV). Additionally, one count per second in GALEX corresponds to reference fluxes of $1.40\times10^{-15}$ and $2.06\times10^{-16}$ erg cm$^2$ s$^{-1}$\AA$^{-1}$ in the FUV and NUV, respectively. Conversion of the given magnitudes to fluxes in erg cm$^2$ s$^{-1}$\AA$^{-1}$ is thus achieved using the following formulae:

\begin{equation}
f_{FUV} = 1.40\times10^{-15} \cdot 10^{\frac{m_{AB,FUV} - 18.82}{-2.5}}
\end{equation}

\begin{equation}
f_{NUV} = 2.06\times10^{-16} \cdot 10^{\frac{m_{AB,NUV} - 20.08}{-2.5}}
\end{equation}

Conversion to frequency units is then simply achieved by multiplying each flux by $\lambda^{2}/c$, where $\lambda$ is the peak wavelength of the band, and can then be directly converted to mJy. 

\subsubsection{\emph{UBV} and 2MASS Magnitudes}
\label{sec:ubv2mass}

The conversion to flux density is simpler for the $UBV$ and 2MASS magnitudes, as their flux density zero points are known and tabulated in the proper units for our purposes. In this case, the flux in Jy for any band \emph{A} is calculated using:

\begin{equation}
f_{A} = f_{A,0} \cdot 10^{\frac{m_{A}}{-2.5}}
\end{equation}

Aiming to facilitate optical source selection of interesting targets in the \emph{Kepler} field, \citet{everett12} completed a \emph{UBV} photometric survey of the field with the NOAO Mosaic-1.1 Wide Field Imager and the WIYN 0.9 m telescope on Kitt Peak. The magnitudes are in the traditional Vega-based Johnson system, and can be converted easily to fluxes using the standard tabulated fluxes at zero magnitude of the \emph{U}, \emph{B}, and \emph{V} bands of 1823, 4130, and 3781 Jy. 

The absolute calibration of the 2MASS survey was performed by \citet{cohen03}. The fluxes at zero magnitude for the $JHK$ filters are 1594, 1024, and 666.8 Jy, respectively.

\begin{figure*}
\begin{center}
\includegraphics[width=\textwidth]{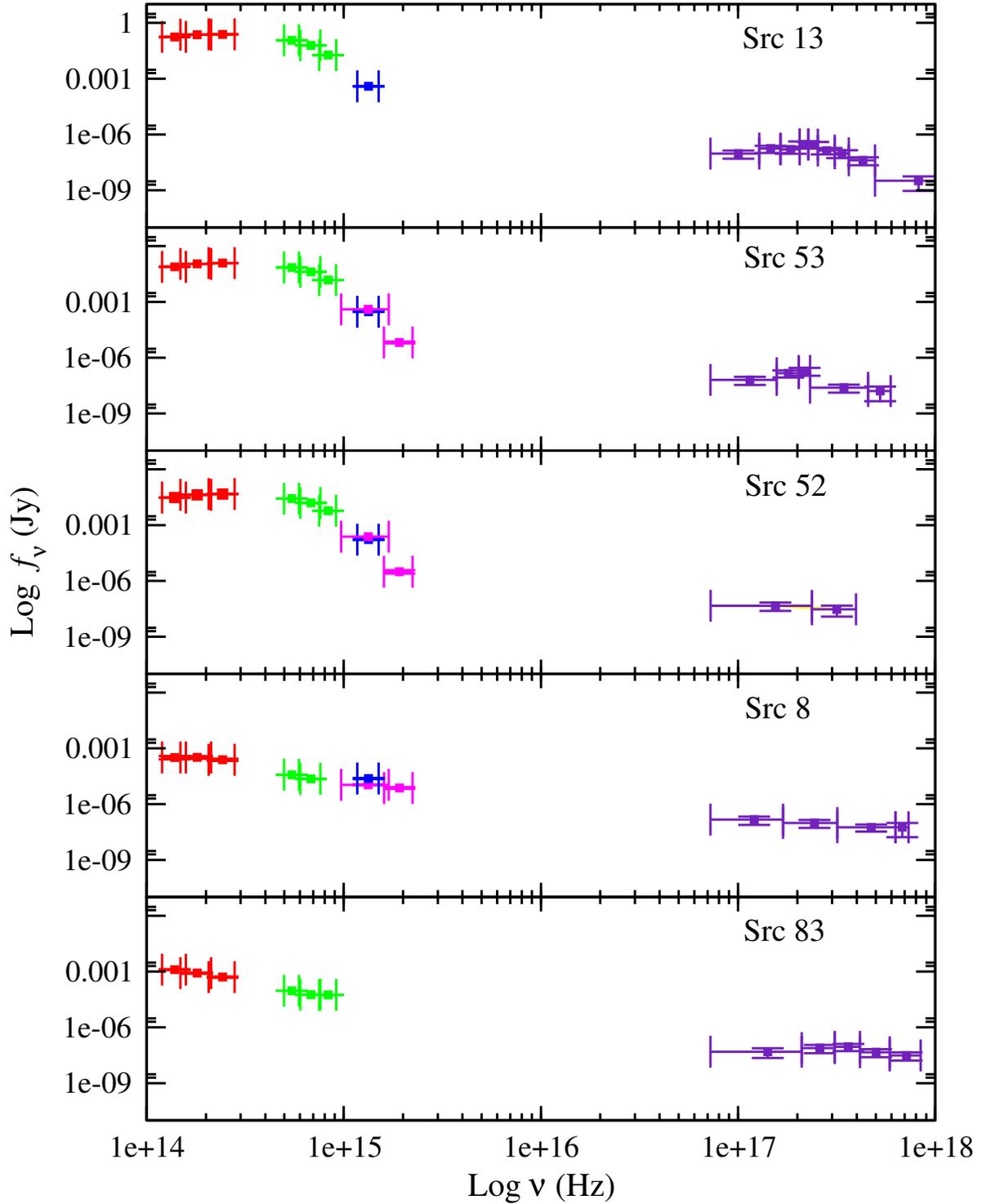}
\figcaption[]{Broadband SEDs for three stellar objects (top three panels) and two AGN (bottom two panels). Red points are the 2MASS JHK colors, green points are UBV fluxes calculated from the \citet{everett12} survey, blue points are UVOT M2 filter fluxes, magenta points are GALEX FUV and NUV fluxes, and purple points are XRT fluxes from this survey. The objects were selected as stellar or AGN based on their log $f_X/f_V$ values (i.e., their position on Figure \ref{fig:optrat}).  
\label{fig:seds}}
\end{center}
\end{figure*}

With all given magnitudes converted to consistent flux units, we constructed the broadband SEDs for objects that have data in at least three wavebands. Plotted in Figure~\ref{fig:seds} are the SEDs of five representative objects from different regions of the $f_X / f_V$ plot; three stellar objects and two AGN. This analysis is only informative for objects with data in 2MASS, the \citet{everett12} UBV survey, and/or GALEX, so we have displayed the SEDs for objects of each type with the most available data points. Note that AGN will necessarily have fewer detections in the ancillary surveys, since they are relatively much fainter on average than stellar sources in the optical and UV. Only two of the sources have GALEX data. The GALEX NUV filter overlaps with the UVOT M2 filter; however, because the observations are widely disparate in time, discrepancies in the flux values might indicate time variability. Additionally, since the 2MASS and UBV measurements are also disparate in time, SED fitting is not appropriate. 

Nonetheless, one can clearly see that the AGN-type SEDs are flatter than the stellar SEDs at lower frequencies, and differ characteristically enough to allow for reliable source classification.

\section{Sample Light Curves}
\label{sec:lcs}

The driving motivation behind this survey is the unprecedented photometric precision of \emph{Kepler} and its application to astrophysical sources beyond exoplanet detection. Over its $\sim$4 year operational lifetime, \emph{Kepler} routinely monitored over 150,000 stars, the overwhelming majority of which show no evidence of transiting exoplanets due to simple geometry. The unmatched photometric quality of the dataset will result in the \emph{Kepler} targets being the best monitored astrophysical sources ever, and ancillary data will help insure they are also the best understood.
There are many types of X-ray bright, optically variable sources both within and beyond the Galaxy. The MAST Kepler Data Search tool allows the user to input coordinates, object names, or KIC ID numbers and obtain all available quarterly light curves for any object. All \emph{Kepler} data is now public, so if an object has been observed, in general one can download and analyze the light curves.

To demonstrate the temporal analyses to be carried out on the survey products, we have selected one representative object from each class (stellar and AGN). The following sections constitute a sample of results to be presented in upcoming papers focusing on the spectral and time-series analysis of each source type. 
\\

\begin{figure}
    \centering
    \includegraphics[width=0.5\textwidth]{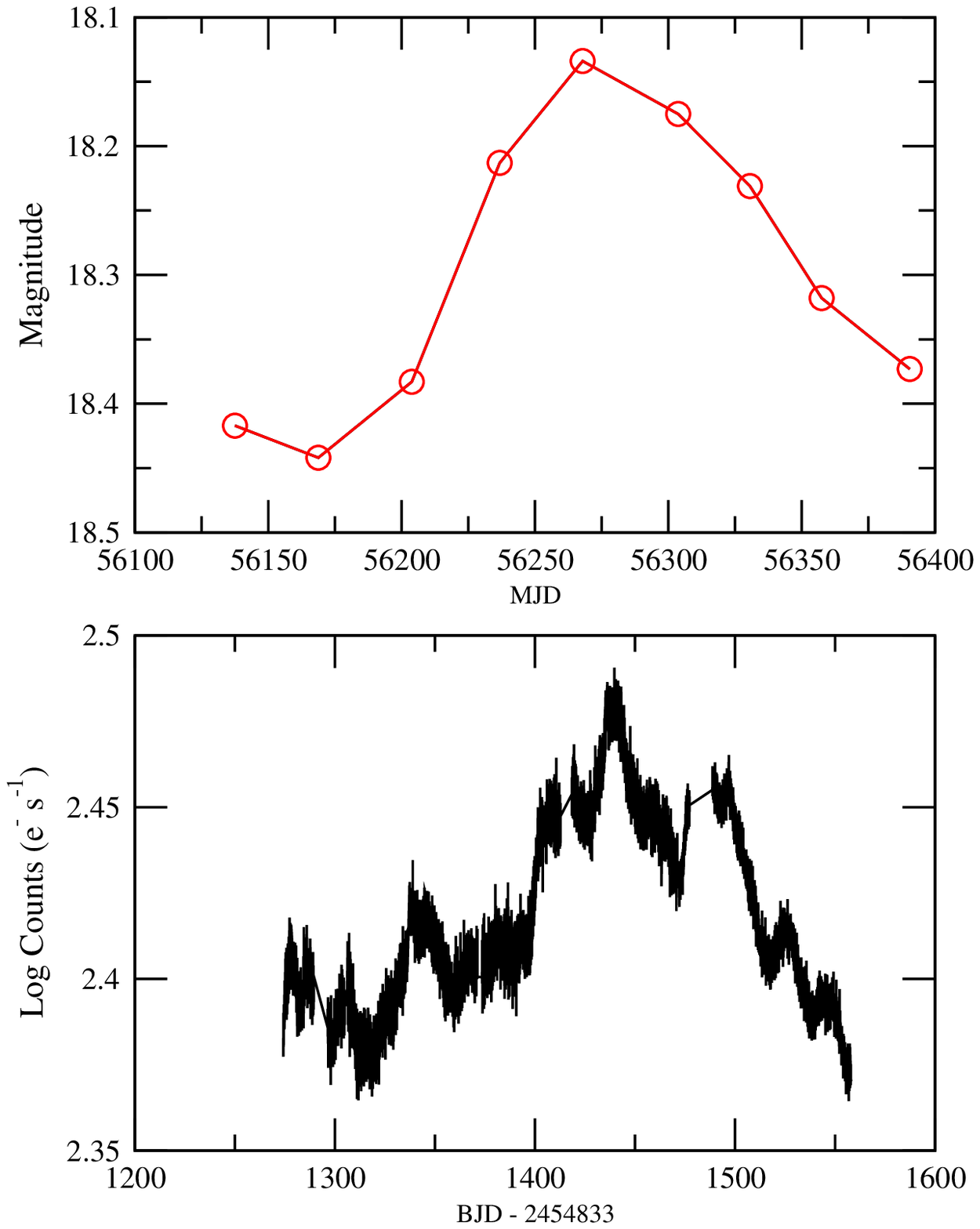}
    \figcaption{The full-frame image (FFI) light curve with 30-day sampling (top panel) and the standard \emph{Kepler} light curve with 30-minute sampling (bottom panel) for KSw~9 (KIC~7175757), a BL~Lac type AGN.  The bottom panel's flux units have been put into a log scale for consistency with the magnitude units in the top panel. Both light curves are the output of our custom pipeline. While the FFI light curve involves different extraction apertures and procedures, as well as entirely different source images, the behavior of the variability is faithfully tracked.
    \label{fig:ffi}}
\end{figure}

\subsection{AGN: Timing Overview}
\label{sec:agnsample}

The discovery of AGN in the \emph{Kepler} FOV was the original primary motivation for this survey. Although a few AGN were known to be in the field, a lack of overlap with large spectroscopic surveys like SDSS prevented the curation of a sizable sample. Additionally, the known AGN in the field were typically selected by techniques known to bias the final sample. X-ray detection is the least-biased AGN selection method, as it is immune to all but the most intense dust obscuration and to the effect of dilution of the optical or infrared colors by host galaxy starlight \citep{mush04}. Our survey is approximately ten times deeper than the ROSAT All-Sky Survey (RASS), which included only a handful of confirmed AGN in this region of the sky. 

The optical emission in an AGN is believed to originate in an accretion disk. The variability of this emission then implies stochastic processes within the disk. There are several candidate processes that theoretically give rise to rapid optical variability, including reprocessing of the variations of the central X-ray source, the orbital dynamics of the gas in the disk itself, spatial viscosity variations, and turbulent thermal processes. A clear understanding of the relationship of AGN parameters to overall variability requires a self-consistent, complete theory of accretion disks including radiative physics; this does not yet exist and there are many hurdles left for such simulations. The approach has therefore been largely empirical, involving observed correlations like the \citet{mchardy04} study showing a relationship between the characteristic timescale of X-ray variability and black hole mass spanning many orders of magnitude, from stellar-mass black hole systems like Cygnus~X-1 to Type 1 Seyfert galaxies. Such observations remain unexplained, but seem to be robust for the 2-10 keV band \citep[e.g.,][]{gonzalez12, ludlam15}. Additional approaches have involved mathematical modeling of observed variability as a damped random walk model \citep{macleod10} and propagating fluctuations that produce flicker noise \citep{lyubarskii97}. 

The \emph{Kepler} data are superior to all previous AGN light curves ever obtained, in both photometric precision and continuous sampling frequency. The proper analysis techniques to analyze these light curves must be carefully constructed to mitigate the systematic errors known to be present in the \emph{Kepler} data, such as the Moir\'{e} pattern drift noise \citep{kolod10} and the inter-quarter discontinuities introduced by spacecraft rolls. 

Orbital timescales for typical AGN black hole masses ($10^6 - 10^9 M_{\odot}$) span a few days to months; timescales which we can probe using the \emph{Kepler} light curves. \citet{edelson14} reported the first detection of an optical timescale break, detected at $\sim5$ days, for one of our X-ray detected previously-known AGN (Zw~229-15). A small number of our confirmed AGN have continuous \emph{Kepler} monitoring (see Table~\ref{t:tab2}, columns with ``Y"). Our original intent was to propose for \emph{Kepler} to monitor all KSwAGS X-ray sources as soon as we had confirmed them; however, only four of the newly discovered sources were being monitored at the time of the failure of the second reaction wheel that put an end to the original \emph{Kepler} mission. For those objects without archived light curves, we are able to construct coarsely sampled light curves with 30-day cadence for the entirety of the 4 year mission using the full-frame images (FFIs): snapshots of the entire FOV downloaded each month to verify pointing calibrations. We have written a customized pipeline to handle the FFI data cubes and produce a light curve for any source that was on silicon, whether or not it was included in the Kepler Input Catalog (KIC). Figure~\ref{fig:ffi} demonstrates that our FFI method faithfully tracks the variability in the archived light curve for KSw~9 (KIC~7175757; BZB~J1848+4245). The displayed data is typical of the AGN light curves in our sample: aperiodic and stochastic, with both short- and long-term variability and with higher amplitudes at longer timescales. 

An upcoming paper in this series will explore in detail the variability of the AGN in our sample, using our custom pipelines for both the FFIs and the archived data. It will also include all of the spectra and measurable X-ray properties of our confirmed AGN. We plan to use a variety of time series analysis techniques to search for characteristic timescales and correlations of variability properties with the measured parameters. Our spectroscopic campaign is currently continuing on the 4.3-meter Discovery Channel Telescope at Lowell Observatory.
\\
\\
\subsection{Stars: Timing Overview}
\label{sec:starexample} 

As described in Section~\ref{sec:starspec}, the Palomar spectra of the KSwAGS survey stars fall into roughly three groups: M dwarfs with strong line emission, normal main sequence spectra, and G through K stars with broad and complex emission structures in their H alpha lines and the Ca~H and K lines. Their $v$sin$i$ measurements cluster around roughly 100 km/sec, implying that, as a group, they are rapid rotators. Rapid rotation and X-ray luminosity are strongly correlated, and such stars are concentrated in two basic stellar populations: young stars, which have not had time to lose their angular momentum, and binary stars, where tidal transfer of orbital momentum maintains the rotational angular momentum.

\begin{figure*}
    \centering
    \includegraphics[width=\textwidth]{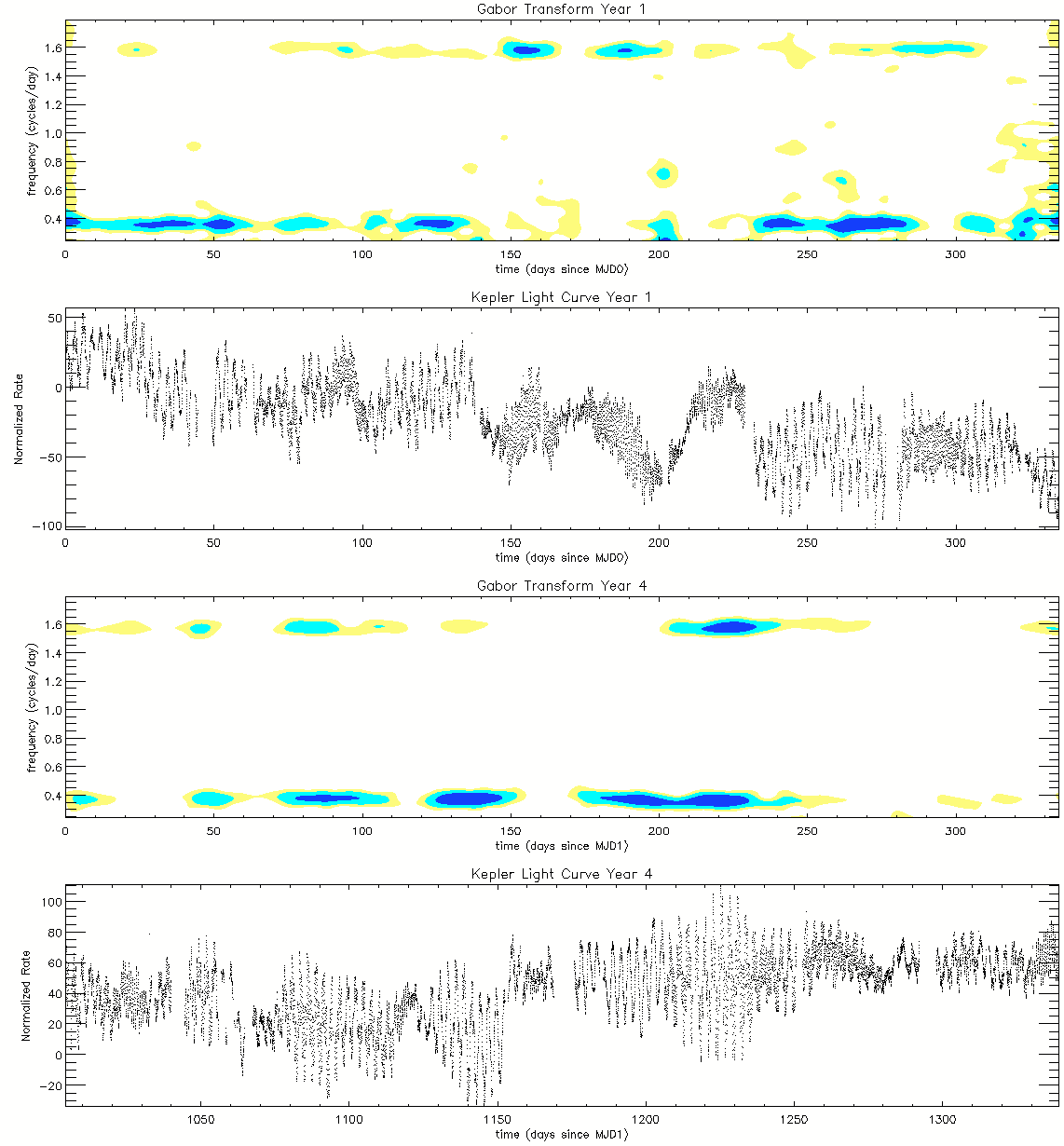}
    \caption{Light curves and dynamical power spectra (also known as Gabor transforms) for the first and fourth years of continuous \emph{Kepler} monitoring of the variable star KSw~47 (KIC~6365080). There are two dominant periods with constant frequency but varying amplitude. Such light curves appear frequently in our sample of X-ray bright, \emph{Kepler}-monitored stars.}
    \label{fig:starsample}
\end{figure*}

Preliminary investigation of the 27 \emph{Kepler} light curves of KSwAGS stellar sources supports this. Fourier analysis of these sources detects highly significant periods in 22 of the 27 objects. Twenty have periods of 10 days or less, and nine of these have periods less than one day. Most show evidence of slow period and amplitude evolution over multiple \emph{Kepler} quarters. Figure~\ref{fig:starsample} shows the light curves and dynamic power spectra over two years for KSw~47 (KIC~6365080). In this example, we note the complex interplay between two fundamental periods at $\sim$2.8 days and $\sim$0.64 days, with constant frequencies but dramatically varying amplitudes. \citet{mcquillan14} include this object in their large sample of rotationally variable stars in the \emph{Kepler} field, and attribute the longer period to rotation; however, they only exclude pulsating stars by making a simple cut in effective temperature, and have likely not excluded all pulsators. The effective temperature in the KIC is given as $T_{eff}=5900~\rm{K}$ from $griz$ photometry; however, our Palomar spectrum suggests an F type star. This is not necessarily unusual: the KIC temperatures may be $\sim100-200$K too cool, as reported by \citet{pinson12}. If the shorter period is indeed due to pulsation, then the frequency and stellar type suggest the object is an RR~Lyrae star, which typically have periods of $0.2-1$~days. Such complex, multi-periodic light curves are typical for a number of KSwAGS targets. Further investigation of \emph{Kepler} photometry will allow us to track the growth, migration, and decay of starspots, differential rotation, activity cycles, and flaring on a wide variety of single and binary stars, offering a unique opportunity to measure magnetic activity cycles for a large sample of late-type stars, which could provide important constraints for developing better stellar dynamo models and a clearer understanding of how they function. The timing properties of the KSwAGS stellar sources will be fully investigated in an upcoming paper in this series.

\section{Continuation in the K2 Fields}
\label{sec:k2}
Since the second reaction wheel failure and the resulting loss of pointing stability in the original FOV, the \emph{Kepler} mission has been modified and repurposed as ``K2". The new mission utilizes solar radiation pressure and the remaining two wheels to maintain pointing for approximately 3 months at a time, in fields aligned along the ecliptic plane \citep{howell14}. The new incarnation has been proven to produce photometric precision on par with that of the nominal mission, according to the K2 photometry status report conducted after the redesign in December 2013. 

We have surveyed K2 Field 4 with \emph{Swift} using the same array of pointings as our original survey described in this paper. We supplemented our successful K2 monitoring proposal with sources from various additional X-ray missions including \emph{ROSAT}, \emph{Chandra}, and XMM-\emph{Newton}, as well as from the Million Quasar Catalog (``Milliquas")\footnote{Milliquas is unpublished but can be accessed at http://quasars.org/milliquas.htm or via the NASA HEASARC at http://heasarc.gsfc.nasa.gov/W3Browse/all/milliquas.html.}. Additionally, we have submitted the pointing coordinates for K2~Fields~8 and 10 to \emph{Swift}; the survey in those fields will begin presently. We expect to continue to publish future versions of the KSwAGS survey as we increase our sample in the new fields; additionally, all future KSwAGS sources will have 30-minute cadence light curves for the full 3 month duration of each field pointing. The K2 campaign covers a wide range of galactic latitudes; the field we have surveyed so far, Field 4, is in the constellation Taurus at an approximate galactic latitude of $b\sim15^{\circ}$. We have communicated with the \emph{Kepler} and K2 Guest Observer (GO) offices to add our X-ray and UV survey products to the KIC and the EPIC (Ecliptic Plane Input Catalog), so that the wider astronomical community can access these data for their own analyses.

\section{Summary}
\label{sec:summary}

This paper has introduced and described the KSwAGS survey, a \emph{Swift} XRT and UVOT survey of four modules of the \emph{Kepler} FOV. The survey discovered 93 total X-ray sources with S/N$\geq$3, with exposure times of $\sim$2000 seconds per pointing. Of these, 60 have counterparts that were observed simultaneously with UVOT (the remaining 33 were not in the corresponding UVOT images of any of our XRT pointings, due to the smaller FOV of the UVOT telescope). The aim of the survey was to obtain X-ray sources that are likely to have optical counterparts with variability of astrophysical interest, especially AGN. The survey also produced a large number of stellar sources, both new and previously known in the literature. In most cases, the KSwAGS survey provides the first X-ray and UV observations for these objects.

Twenty-three of the source counterparts had optical classifications in the literature. We obtained optical spectra at Palomar for an additional 30 sources. Of these 53 total sources with certain identification, 13 are AGN and 40 are stars. In the absence of spectra, we demonstrate that most of the survey sources can be categorized as either stellar or extragalactic using the flux ratio $f_X/f_V$ or broadband SED shape.

As a sample of the future analysis to be carried out on the KSwAGS sources, we have shown example light curves  and dynamical power spectra of a typical star from our survey, and both types of light curves (FFI and standard \emph{Kepler}) for an AGN.  

The KSwAGS survey has identified numerous new X-ray sources in the original \emph{Kepler} field which can be followed up either using archived light curves from the KIC database, or by constructing 30-day cadence light curves using the FFIs, which can be done for any source in the survey regardless of whether it was monitored by \emph{Kepler}. This opens up a new phase space of X-ray and UV bright targets with high-quality optical time sampling. The survey is currently continuing in Fields~4, 8 and 10 of the K2 mission, the new extension of \emph{Kepler} to the ecliptic plane, and will continue to yield a rich crop of X-ray bright, optically variable targets for monitoring with the most exquisite photometer of our time.

\acknowledgments

We would like to thank the referee for a helpful report which improved the manuscript. We acknowledge Trisha Doyle for her assistance during the Palomar observing run. We also acknowledge the extremely helpful and accommodating staff at Palomar Observatory. The GALEX data for many of our KSwAGS sources was obtained thanks to A. Brown's GALEX GO programs GI4-056 and GI5-055, a UV survey specifically designed to locate active stars in the \emph{Kepler} field. This publication makes use of data products from the Wide-field Infrared Survey Explorer, which is a joint project of the University of California, Los Angeles, and the Jet Propulsion Laboratory/California Institute of Technology, funded by the National Aeronautics and Space Administration. KLS is grateful for support from the NASA Earth and Space Sciences Fellowship (NESSF), which enabled the majority of this work.

\newpage

 \renewcommand{\tabcolsep}{2.6pt}
 \footnotesize
 \begin{longtable}{lccccccccccc}
 \caption{\emph{Swift} XRT and UVOT Sources in the \emph{Kepler} FOV} \label{tab:coords} \\
 \hline \hline \\[-2ex]
 \multicolumn{1}{c}{Src} &
 \multicolumn{1}{c}{X-ray RA} &
 \multicolumn{1}{c}{X-ray Dec} &
 \multicolumn{1}{c}{Region} &
 \multicolumn{1}{c}{Pointing} &
 \multicolumn{1}{c}{Ct Rate} &
 \multicolumn{1}{c}{Exp Time (s)} &
 \multicolumn{1}{c}{S/N} &
 \multicolumn{1}{c}{UV RA} &
 \multicolumn{1}{c}{UV Dec} &
 \multicolumn{1}{c}{UVOT Flux (mJy)} &
 \multicolumn{1}{c}{Notes} \\[0.5ex] \hline
 \\[-1.8ex]
 \endfirsthead
 
 \multicolumn{11}{c}{{\tablename} \thetable{ } -- Continued} \\[0.5ex]
 \hline \hline \\[-2ex]
 \multicolumn{1}{c}{Src} &
 \multicolumn{1}{c}{X-ray RA} &
 \multicolumn{1}{c}{X-ray Dec} &
 \multicolumn{1}{c}{Region} &
 \multicolumn{1}{c}{Pointing} &
 \multicolumn{1}{c}{Ct Rate} &
 \multicolumn{1}{c}{Exp Time (s)} &
 \multicolumn{1}{c}{S/N} &
 \multicolumn{1}{c}{UV RA} &
 \multicolumn{1}{c}{UV Dec} &
 \multicolumn{1}{c}{UVOT Flux (mJy)} &
 \multicolumn{1}{c}{Notes}\\[0.5ex] \hline
 \\[-1.8ex]
 \endhead
 
 \hline

  \hline
 \endfoot
 
 \\[-1.8ex] \hline \hline
 \endlastfoot
 
1	&	280.9950	&	43.4720	&	1	&	12	&	0.0105	&	1829.18	&	3.839	&	280.9955	&	43.4743	&	8.075	&		\\
2	&	281.0685	&	43.6865	&	1	&	13	&	0.00691	&	1770.9	&	3.103	&	281.0698	&	43.686	&	0.004	&		\\
3	&	281.1287	&	43.2791	&	1	&	11	&	0.00803	&	1617.21	&	3.105	&	281.1273	&	43.2803	&	0.041	&		\\
4	&	281.2845	&	42.6883	&	1	&	16	&	0.00793	&	1722.52	&	3.146	&	281.2855	&	42.688	&	0.013	&		\\
5	&	281.5498	&	42.6175	&	1	&	16	&	0.0187	&	1720.15	&	4.963	&	281.5518	&	42.6183	&	0.007	&		\\
6	&	281.7771	&	43.6767	&	1	&	28	&	0.0168	&	1674.88	&	4.207	&		&		&		&	No UVOT data	\\
7	&	281.8198	&	42.3348	&	1	&	24	&	0.011	&	1601.88	&	3.756	&	281.8209	&	42.3336	&	0.018	&		\\
8	&	282.1510	&	44.8243	&	1	&	32	&	0.0129	&	1668.11	&	4.184	&	282.1502	&	44.8241	&	0.024	&		\\
9	&	282.1965	&	42.7612	&	1	&	35	&	0.228	&	1589.43	&	18.11	&	282.1963	&	42.7609	&	0.031	&		\\
10	&	282.2645	&	43.7389	&	1	&	37	&	0.0125	&	1666.28	&	3.998	&		&		&		&	No UVOT data	\\
11	&	282.3341	&	43.7350	&	1	&	37	&	0.0137	&	1888.35	&	4.43	&		&		&		&	No UVOT data	\\
12	&	282.3970	&	43.7171	&	1	&	37	&	0.00752	&	1920.81	&	3.25	&		&		&		&	No UVOT data	\\
13	&	282.6576	&	43.4456	&	1	&	43	&	0.0344	&	1940.96	&	7.538	&	282.6572	&	43.4447	&	0.390	&		\\
14	&	282.7376	&	42.9820	&	1	&	42	&	0.0205	&	1702.46	&	5.138	&	282.737	&	42.9819	&	1.130	&		\\
15	&	283.1236	&	43.6753	&	1	&	50	&	0.00853	&	1873.03	&	3.381	&	283.1242	&	43.6752	&	0.044	&		\\
16	&	283.4132	&	43.1623	&	1	&	52	&	0.00747	&	1933.13	&	3.158	&		&		&		&	Out of FOV	\\
17	&	283.4805	&	43.4608	&	1	&	53	&	0.0121	&	1427.32	&	3.08	&		&		&		&	Out of FOV	\\
18	&	283.5507	&	43.2023	&	1	&	53	&	0.0094	&	1702.04	&	3.388	&	283.5496	&	43.2029	&	0.071	&		\\
19	&	286.3436	&	43.4668	&	2	&	8	&	0.0104	&	1440.55	&	3.024	&	286.3436	&	43.4675	&	0.115	&		\\
20	&	286.3582	&	42.4609	&	2	&	10	&	0.257	&	1096.97	&	15.7	&	286.3581	&	42.4611	&	0.451	&		\\
21	&	286.5829	&	42.5433	&	2	&	11	&	0.0112	&	1767.65	&	3.69	&		&		&		&	Out of FOV	\\
22	&	286.8068	&	41.5243	&	2	&	20	&	0.0796	&	1274.54	&	8.031	&		&		&		&	Out of FOV	\\
23	&	286.8176	&	44.0189	&	2	&	16	&	0.122	&	1573.93	&	11.24	&		&		&		&	Out of FOV	\\
24	&	287.0422	&	43.7961	&	2	&	24	&	0.0106	&	1863.42	&	3.92	&	287.0419	&	43.7962	&	0.009	&		\\
25	&	287.1229	&	42.3448	&	2	&	25	&	0.0118	&	1611.42	&	3.761	&	287.1238	&	42.344	&	0.123	&		\\
26	&	287.2791	&	44.1560	&	2	&	26	&	0.00793	&	1714.99	&	3.133	&		&		&		&	No UV CP	\\
27	&	287.2975	&	41.5352	&	2	&	20	&	0.0184	&	1598.75	&	4.455	&		&		&		&	Out of FOV	\\
28	&	287.7060	&	42.9274	&	2	&	36	&	0.0121	&	3357.91	&	5.152	&	287.7059	&	42.9268	&	3.410	&		\\
29	&	287.7355	&	43.5900	&	2	&	40	&	0.00531	&	3473.52	&	3.79	&	287.7359	&	43.5907	&	0.044	&		\\
30	&	287.8162	&	44.1634	&	2	&	35	&	0.00492	&	3035.28	&	3.361	&		&		&		&	Out of FOV	\\
31	&	287.8831	&	42.8539	&	2	&	36	&	0.00565	&	2787.61	&	3.06	&	287.8829	&	42.856	&	0.008	&		\\
32	&	287.9496	&	42.0781	&	2	&	39	&	0.00375	&	3131.29	&	2.998	&		&		&		&	Out of FOV	\\
33	&	287.9973	&	41.8501	&	2	&	39	&	0.0078	&	3349.98	&	4.565	&	287.9983	&	41.8504	&	0.020	&		\\
34	&	288.0440	&	43.1298	&	2	&	44	&	0.00906	&	2662.11	&	4.21	&	288.0438	&	43.1288	&	0.054	&		\\
35	&	288.2545	&	42.2036	&	2	&	47	&	0.0119	&	1629.74	&	3.688	&	288.2551	&	42.2034	&	0.120	&		\\
36	&	288.3293	&	42.4670	&	2	&	42	&	0.0303	&	423.119	&	3.138	&		&		&		&	Out of FOV	\\
37	&	288.5651	&	42.0827	&	2	&	47	&	0.0344	&	1629.74	&	6.477	&		&		&		&	Out of FOV	\\
38	&	288.5717	&	42.6082	&	2	&	49	&	0.0248	&	1649.34	&	5.6	&	288.5718	&	42.6089	&	0.046	&		\\
39	&	288.6929	&	42.3920	&	2	&	52	&	0.0134	&	1709.98	&	4.119	&	288.6932	&	42.3918	&	0.080	&		\\
40	&	288.7612	&	43.3231	&	2	&	51	&	0.0155	&	1714.79	&	4.496	&	288.7613	&	43.323	&	0.060	&		\\
41	&	291.1622	&	42.7210	&	3	&	3	&	0.00728	&	1767.81	&	3.078	&		&		&		&	Out of FOV	\\
42	&	291.3922	&	41.7268	&	3	&	4	&	0.0108	&	1711.13	&	3.846	&	291.392	&	41.7271	&	0.012	&		\\
43	&	291.5486	&	42.7656	&	3	&	3	&	0.0148	&	1379.1	&	3.989	&		&		&		&	Out of FOV	\\
44	&	291.6264	&	41.5510	&	3	&	8	&	0.0841	&	1115.95	&	9.203	&		&		&		&	Out of FOV	\\
45	&	291.6717	&	42.1556	&	3	&	6	&	0.00688	&	1751.02	&	3.12	&		&		&		&	Out of FOV	\\
46	&	292.0210	&	42.0779	&	3	&	10	&	0.0066	&	1867.94	&	3.01	&		&		&		&	No UV CP	\\
47	&	292.1626	&	41.7436	&	3	&	16	&	0.0101	&	1878.94	&	3.611	&		&		&		&	Out of FOV	\\
48	&	292.1953	&	42.7736	&	3	&	15	&	0.00715	&	1836.44	&	3.109	&	292.1957	&	42.7734	&	0.009	&		\\
49	&	292.2361	&	43.0941	&	3	&	17	&	0.0261	&	1824.52	&	6.291	&	292.235	&	43.0936	&	0.204	&		\\
50	&	292.3142	&	42.6770	&	3	&	15	&	0.0116	&	1893.66	&	3.99	&	292.313	&	42.6764	&	0.027	&		\\
51	&	292.4233	&	41.2553	&	3	&	22	&	0.00643	&	1997.72	&	3.101	&		&		&		&	No UV CP	\\
52	&	292.5009	&	42.2127	&	3	&	19	&	0.00669	&	1852.89	&	2.999	&	292.502	&	42.2138	&	0.165	&		\\
53	&	292.6304	&	42.8299	&	3	&	23	&	0.0148	&	1948.25	&	4.669	&	292.6299	&	42.8293	&	0.293	&		\\
54	&	292.6719	&	43.0381	&	3	&	25	&	0.00965	&	2027.3	&	3.918	&		&		&		&	Out of FOV	\\
55	&	292.8015	&	43.2247	&	3	&	28	&	0.0331	&	1885.49	&	6.909	&	292.8021	&	43.2243	&	0.465	&		\\
56	&	293.0162	&	41.0456	&	3	&	37	&	0.0176	&	1095.07	&	4.012	&		&		&		&	Out of FOV	\\
57	&	293.1699	&	42.8071	&	3	&	33	&	0.0136	&	1785.2	&	4.055	&	293.1702	&	42.8074	&	0.048	&		\\
58	&	293.1876	&	41.0621	&	3	&	37	&	0.00863	&	1573.6	&	3.3	&		&		&		&	No UV CP	\\
59	&	293.2583	&	41.6922	&	3	&	42	&	0.00461	&	3048.38	&	3.052	&	293.2581	&	41.6923	&	0.016	&		\\
60	&	293.3043	&	43.1656	&	3	&	43	&	0.0141	&	1565.68	&	3.917	&	293.3041	&	43.1653	&	0.009	&		\\
61	&	293.4492	&	41.1218	&	3	&	37	&	0.0141	&	1433.14	&	4.118	&		&		&		&	Out of FOV	\\
62	&	293.5059	&	41.3211	&	3	&	46	&	0.0248	&	2363.78	&	6.557	&	293.5067	&	41.3207	&	0.006	&		\\
63	&	293.6725	&	42.4138	&	3	&	45	&	0.0213	&	1475.13	&	4.429	&		&		&		&	Out of FOV	\\
64	&	293.8478	&	41.2915	&	3	&	46	&	0.00992	&	2484.73	&	4.114	&		&		&		&	Out of FOV	\\
65	&	293.9017	&	42.8962	&	3	&	47	&	0.0111	&	2231.23	&	4.395	&	293.9013	&	42.896	&	0.040	&		\\
66	&	293.9578	&	42.3660	&	3	&	51	&	0.00478	&	2707.55	&	3.142	&	293.9579	&	42.366	&	0.092	&		\\
67	&	294.0024	&	41.9892	&	3	&	49	&	0.00803	&	2040.93	&	3.407	&	294.0035	&	41.9896	&	0.015	&		\\
68	&	294.1042	&	42.4103	&	3	&	51	&	0.0115	&	2672.43	&	4.812	&	294.1045	&	42.4106	&	0.039	&		\\
69	&	294.1987	&	41.7921	&	3	&	53	&	0.00566	&	2544.91	&	3.224	&	294.2041	&	41.791	&	0.445	&		\\
70	&	294.3325	&	41.6892	&	3	&	53	&	0.00658	&	2544.91	&	3.11	&	294.3319	&	41.6852	&	2.700	&		\\
71	&	294.3559	&	41.7774	&	3	&	53	&	0.0204	&	2544.91	&	6.22	&	294.3562	&	41.7774	&	0.098	&		\\
72	&	294.4295	&	41.7054	&	3	&	55	&	0.00585	&	3116.57	&	3.366	&	294.43	&	41.7055	&	0.015	&		\\
73	&	296.4893	&	41.7504	&	4	&	3	&	0.0104	&	1948.69	&	4.04	&		&		&		&	Out of FOV	\\
74	&	296.5844	&	40.7630	&	4	&	4	&	0.0102	&	2099.18	&	4.029	&		&		&		&	No UV CP	\\
75	&	296.8341	&	40.9947	&	4	&	5	&	0.00906	&	1913.95	&	3.571	&	296.8311	&	40.9946	&	91.958	&		\\
76	&	296.9863	&	41.5419	&	4	&	8	&	0.0118	&	1781.26	&	3.856	&	296.9858	&	41.541	&	0.020	&		\\
77	&	297.0958	&	41.3399	&	4	&	6	&	0.0162	&	1883.6	&	4.416	&		&		&		&	Out of FOV	\\
78	&	297.1825	&	39.9171	&	4	&	11	&	0.0132	&	1977.29	&	4.41	&	297.1835	&	39.9169	&	108.980	&		\\
79	&	297.3499	&	41.5886	&	4	&	17	&	0.0959	&	1822.85	&	11.45	&	297.3492	&	41.5892	&	0.021	&		\\
80	&	297.3761	&	41.6051	&	4	&	17	&	0.00916	&	1871.98	&	3.457	&	297.3765	&	41.6054	&	0.009	&		\\
81	&	297.3857	&	39.6107	&	4	&	15	&	0.00652	&	1896.59	&	3.008	&	297.3849	&	39.6103	&	0.264	&		\\
82	&	297.5560	&	42.1157	&	4	&	20	&	0.012	&	1641.57	&	3.661	&		&		&		&	Out of FOV	\\
83	&	297.7414	&	40.9773	&	4	&	19	&	0.0117	&	2939.39	&	5.362	&		&		&		&	Out of FOV	\\
84	&	297.8264	&	41.3590	&	4	&	29	&	0.0259	&	1515.32	&	5.037	&		&		&		&	Out of FOV	\\
85	&	297.8535	&	40.7351	&	4	&	26	&	0.00933	&	1792.64	&	3.514	&	297.8532	&	40.7355	&	21.210	&		\\
86	&	297.9148	&	40.1621	&	4	&	28	&	0.00922	&	1731.57	&	3.493	&		&		&		&	No UV CP	\\
87	&	297.9698	&	41.6411	&	4	&	32	&	0.00926	&	1558.74	&	3.061	&		&		&		&	Out of FOV	\\
88	&	298.4782	&	40.8922	&	4	&	43	&	0.00871	&	1525.06	&	3.182	&		&		&		&	Out of FOV	\\
89	&	298.6474	&	41.4640	&	4	&	39	&	0.00835	&	1703.89	&	3.221	&	298.647	&	41.4643	&	2.831	&		\\
90	&	298.7180	&	41.9749	&	4	&	44	&	0.00662	&	1817.06	&	3.02	&		&		&		&	No UV CP	\\
91	&	298.7625	&	40.9228	&	4	&	43	&	0.013	&	1710.56	&	4.163	&	298.7625	&	40.9219	&	0.311	&		\\
92	&	298.9334	&	41.9845	&	4	&	44	&	0.0301	&	1817.79	&	6.659	&		&		&		&	No UV CP	\\
93	&	299.8686	&	40.7339	&	4	&	55	&	0.607	&	1634.21	&	28.82	&		&		&		&	No UV CP	\\
 
 \end{longtable}
 \newpage

\LongTables
 \begin{deluxetable}{cccccccccccccccc}
 \tablewidth{0pt}
 \tabletypesize{\tiny}
 \label{tab:list}
 \tablecaption{Broadband fluxes and properties of \emph{Swift} sources \label{t:tab2}}
 \tablehead{
 \colhead{Src} &
 \colhead{KIC ID} &
 \colhead{Kep} &
 \colhead{\emph{GALEX}} &
 \colhead{\emph{GALEX}} &
 \colhead{\emph{U}} &
 \colhead{\emph{B}} &
 \colhead{\emph{V}} &
 \colhead{\emph{J}} &
 \colhead{\emph{H}} &
 \colhead{\emph{K}} &
 \colhead{log } &
 \colhead{Light} &
 \colhead{Query} &
 \colhead{Angular} &
 \colhead{ID} \\
 \colhead{ } &
 \colhead{ } &
 \colhead{ Mag} &
 \colhead{FUV} &
 \colhead{NUV} &
 \colhead{ } &
 \colhead{ } &
 \colhead{ } &
 \colhead{ } &
 \colhead{ } &
 \colhead{ } &
 \colhead{$f_X/f_V$ } &
 \colhead{Curve?} &
 \colhead{Coords} &
 \colhead{Sep. (\arcsec)} &
 \colhead{ }}

 \startdata 
 
1	&	7730305	&	9.32	&		&		&	180.96	&	414.91	&		&	735.34	&	572.66	&	389.7	&		&	Y	&	UV	&	2.58	&	Binary Star	\\
2*	&	7868547	&	18.94	&		&		&		&		&		&		&		&		&		&	Y	&	UV	&	0.96	&		\\
3	&	7582708	&	19.18	&	0.019	&	0.021	&		&		&		&		&		&		&		&	Y	&	UV	&	0.3	&		\\
4	&	7091410	&	20.4	&		&		&		&	0.04	&	0.05	&		&		&		&	0.4	&		&	UV	&	0.6	&		\\
6	&	7869590	&	10.87	&	0.008	&	0.129	&	6.46	&	39.26	&	98.81	&	385.56	&	422.57	&	319.99	&	-2.61	&	Y	&	X	&	2.94	&	$\delta$ Cepheid$^{1,2}$	\\
7	&	6837514	&	19.67	&		&		&		&		&		&		&		&		&		&		&	UV	&	0.6	&		\\
8	&	8669504	&	18.95	&	0.007	&	0.011	&		&	0.02	&	0.04	&	0.24	&	0.33	&	0.32	&	0.7	&		&	UV	&	1.14	&		\\
9	&	7175757	&	18.13	&		&	0.036	&	0.07	&	0.08	&	0.1	&		&		&		&	1.54	&	Y	&	UV	&	0.3	&	BL Lac$^{3,4}$	\\
12	&	7939256	&	18.37	&	0.021	&	0.037	&	0.1	&	0.11	&	0.12	&		&		&		&	-0.02	&		&	X	&	3.6	&		\\
13	&	7732964	&	10.95	&		&		&	18.59	&	61.26	&	115.14	&	238.17	&	228.61	&	173.27	&	-2.36	&	Y	&	UV	&	1.8	&	Variable$^2$	\\
14*	&	7339343	&	11.52	&		&		&	18.72	&	49.02	&		&	116.65	&	94.08	&	64.14	&		&	Y	&	UV	&	1.44	&	Puls. Var$^5$	\\
15	&	7871931	&	18.86	&	0.016	&	0.035	&	0.06	&	0.07	&	0.07	&		&		&		&	0.22	&		&	UV	&	0.66	&		\\
16	&	7505473	&	16.22	&		&	2.669	&		&	146	&	573.86	&	30.99	&	33.42	&	29.07	&	-0.72	&	Y	&	X	&	4.92	&		\\
18	&	7587184	&	18.81	&		&	0.013	&	0.04	&		&		&	0.23	&	0.38	&	0.66	&		&		&	UV	&	0.6	&		\\
19	&	7739728	&	12.5	&		&	0.139	&	4.48	&	14.4	&	27.29	&	63.46	&	64.49	&	46.9	&	-2.26	&	Y	&	UV	&	0.24	&	Variable$^2$	\\
20*	&	6932990	&	11.13	&		&		&		&		&		&	4.22	&	5.9	&	0.32	&		&	Y	&	UV	&	0.3	&	Zw 229 (Sy1)$^{6,7}$	\\
22	&	6190679	&	9.03	&		&		&	41.27	&	225.71	&		&	1934.14	&	2035.64	&	1542.87	&		&	Y	&	X	&	0.78	&	K Star$^8$	\\
23	&	8153411	&	12.64	&		&		&	4.04	&	3.46	&	3.13	&	0.72	&	0.74	&	0.56	&	-0.25	&	Y	&	X	&	0.78	&	MV Lyr$^9$ 	\\
24	&	7948154	&	14.14	&		&		&	0.83	&	3.5	&	7.58	&	25.88	&	28.44	&	20.91	&	-1.69	&		&	UV	&	0.36	&		\\
25	&	6849023	&	18.32	&	0.041	&	0.104	&	0.13	&	0.18	&	0.2	&	0.23	&	0.36	&	0.51	&	-0.06	&		&	UV	&	0.54	&		\\
26*	&	8222218	&	20.27	&		&		&		&		&		&		&		&		&		&		&	X	&	1.56	&		\\
27	&	6191857	&	16.52	&		&		&	0.17	&	0.27	&	0.47	&	2.13	&	2.68	&	3.34	&	-0.25	&		&	X	&	0.96	&		\\
28	&	7350496	&	9.326	&	0.042	&	4.014	&	136.27	&		&		&	875.97	&	733.67	&	507.61	&		&	Y	&	UV	&	1.2	&	G Star	\\
29	&	7811562	&	18.86	&		&		&	0.03	&	0.05	&	0.06	&		&		&		&	0.13	&		&	UV	&	0.3	&		\\
30	&	8223265	&	19.13	&		&		&	0.04	&	0.05	&	0.05	&		&		&		&	0.16	&		&	X	&	2.28	&		\\
31	&	7270227	&	16.43	&		&		&	0.23	&	0.52	&	0.85	&	1.22	&	1.2	&	0.79	&	-1.01	&		&	UV	&	0.78	&		\\
32	&	6594085	&	18.47	&		&		&		&	0.05	&	0.05	&		&		&		&		&		&	X	&	4.38	&		\\
33	&	6431946	&	19.36	&		&		&		&		&		&		&		&		&		&		&	UV	&	0.12	&		\\
34	&	7516296	&	18.73	&	0.015	&		&	0.05	&	0.05	&	0.07	&		&		&		&	0.31	&		&	UV	&	0.24	&		\\
35*	&	6766476	&	17.98	&		&		&	0.13	&	0.14	&	0.17	&		&		&		&		&		&	UV	&	0.12	&		\\
37*	&	6595746	&	19.48	&		&		&		&		&		&	0.83	&	1.01	&	1.45	&		&		&	X	&	1.74	&		\\
38	&	7107762	&	13	&		&		&	3.84	&	10.44	&	18.33	&	36.79	&	35.08	&	25.12	&	-1.71	&	Y	&	UV	&	0.54	&		\\
39*	&	6853073	&	18.27	&		&		&	0.12	&	0.16	&	0.18	&	0.23	&	0.32	&	0.69	&	0.04	&		&	UV	&	0.18	&		\\
40	&	7674095	&	18.09	&		&		&	0.07	&	0.06	&	0.06	&		&		&		&	0.56	&		&	UV	&	0.3	&		\\
41*	&	7198225	&	18.89	&		&		&		&		&	0.05	&		&		&		&	0.38	&		&	X	&	2.76	&		\\
42	&	6362752	&	19.02	&		&		&		&		&	0.03	&		&		&		&	0.66	&		&	UV	&	0.42	&		\\
43	&	7199582	&	19.52	&		&		&		&	0.04	&	0.04	&		&		&		&	0.69	&		&	X	&	4.08	&		\\
45*	&	6691018	&		&		&		&		&		&		&		&		&		&		&		&	X	&	2.28	&		\\
46	&	6606776	&	19.11	&		&		&		&		&		&		&		&		&		&		&	X	&	3.72	&		\\
47	&	6365080	&	11.45	&		&		&	29.81	&	63.79	&	92.73	&	112.64	&	94.08	&	66.55	&	-2.8	&	Y	&	X	&	1.8	&	Rot. Var.$^5$ 	\\
48	&	7201595	&	18.46	&		&	0.004	&		&		&		&		&		&		&		&		&	UV	&	0.6	&		\\
49	&	7446357	&	15.8	&		&		&	0.7	&	0.47	&	0.44	&	0.57	&	0.44	&	0.46	&	-0.06	&	Y	&	UV	&	0.72	&	V1504 Cyg$^{10}$	\\
50*	&	7119467	&	18.44	&		&		&		&		&		&		&		&		&		&		&	UV	&	2.46	&		\\
52	&	6779613	&	12.62	&	0.003	&	0.238	&	5.91	&	15.57	&	26.55	&	46.83	&	42.73	&	30.22	&	-2.44	&	Y	&	UV	&	0.24	&	$\gamma$ Doradus$^5$ 	\\
53	&	7284688	&	11.23	&	0.007	&	0.395	&	14.97	&	41	&	70.28	&	121.25	&	109.22	&	76.48	&	-2.51	&	Y	&	UV	&	0.36	&	Ecl. Binary$^{5,8,11,12}$	\\
54	&	7447756	&	7.27	&		&		&	53.46	&		&		&	16191.55	&	24791.34	&	24050.85	&		&		&	X	&	1.5	&	M Star 	\\
55*	&	7610713	&	16.74	&		&		&		&		&		&	1.01	&	1.3	&	1.94	&		&	Y	&	UV	&	0.72	&		\\
56	&	5794742	&	17.65	&	0.027	&	0.051	&	0.11	&	0.12	&	0.23	&	1.04	&	1.56	&	1.76	&	0.05	&		&	X	&	1.68	&		\\
57	&	7286410	&	13.05	&		&		&	2.17	&	8.32	&	16.98	&	48.09	&	51.37	&	37.29	&	-1.93	&	Y	&	UV	&	0.06	&	Variable$^{5,8,13}$	\\
59	&	6289488	&	18.4	&		&		&		&		&		&		&		&		&		&		&	UV	&	0.6	&		\\
60	&	7532798	&	19.46	&		&		&	0.05	&	0.07	&	0.09	&		&		&		&	0.38	&		&	UV	&	1.98	&		\\
61	&	5881515	&	19.09	&		&		&		&		&		&		&		&		&		&		&	X	&	1.44	&		\\
62*	&	6047927	&	18.67	&		&		&		&		&		&	0.17	&	0.41	&	0.56	&		&		&	UV	&	1.38	&		\\
64	&	5966921	&	14.49	&		&		&	0.13	&	0.83	&	2.44	&	23.49	&	28.89	&	22.32	&	-1.23	&	Y	&	X	&	4.02	&		\\
65	&	7288925	&	19.06	&		&		&	0.07	&	0.09	&	0.1	&		&		&		&	0.21	&		&	UV	&	0.6	&		\\
66	&	6870455	&	7.66	&		&		&	36.44	&		&		&	11420.58	&	18702.51	&	14870.67	&		&		&	UV	&	0.48	&	K Star	\\
67	&	6529378	&	13.82	&		&		&	1.07	&	4.7	&	10.93	&	38.7	&	42.73	&	32.5	&	-1.97	&	Y	&	UV	&	0.66	&		\\
68	&	6956279	&	19.12	&		&		&	0.04	&	0.05	&	0.06	&		&		&		&	0.49	&		&	UV	&	0.18	&		\\
69	&	6371741	&	14.5	&		&		&	2.13	&		&		&	4.29	&	2.85	&	1.95	&		&	Y	&	UV	&	0.66	&		\\
70	&	6293269	&	18.13	&		&		&		&	0	&	84.65	&	33.15	&	38.15	&	30.72	&	0.05	&		&	X	&	1.08	&		\\
71	&	6372268	&	11.44	&		&		&	6.15	&	27.67	&	63.16	&	229.56	&	260.56	&	189.12	&	-2.33	&	Y	&	UV	&	0.72	&		\\
72*	&	6372529	&	19.05	&		&		&		&		&		&		&		&		&		&		&	UV	&	0.54	&		\\
73	&	6380580	&	11.38	&		&		&	24.77	&	57.86	&	93.59	&	133.32	&	115.96	&	80.6	&	-2.79	&	Y	&	X	&	1.08	&	$\gamma$ Doradus$^5$ 	\\
75	&	5724440	&	7.874	&		&		&		&		&		&	1808.37	&	1259.8	&	847.09	&		&	Y	&	UV	&	3.72	&	$\delta$ Scuti$^{14,15}$ 	\\
76	&	6224104	&	15.16	&		&		&	0.66	&	1.67	&	2.7	&	4.07	&	3.43	&	2.43	&	-1.2	&		&	UV	&	0.18	&		\\
77*	&	6062112	&	19.74	&		&		&		&		&		&		&		&		&		&		&	X	&	2.46	&		\\
78	&	4857678	&	7.01	&		&		&		&		&		&	5139.12	&	3759.24	&	2572.35	&		&	Y	&	UV	&	0.96	&	F Star$^{15}$	\\
79	&	6225816	&	10.34	&		&		&		&		&		&	671.26	&	709.09	&	538.93	&		&	Y	&	UV	&	2.76	&	Rot. Var.$^5$ 	\\
80	&	6305971	&	19.47	&		&		&		&		&	0.05	&		&		&		&	0.4	&		&	UV	&	0.36	&		\\
81	&	4585976	&	12.23	&		&		&	9.03	&	23.5	&	39.45	&	65.24	&	52.42	&	37.77	&	-2.62	&	Y	&	UV	&	0.036	&	$\gamma$ Doradus$^5$ 	\\
82*	&	6714686	&	16.01	&		&		&		&		&		&	0.68	&	0.9	&	1.06	&		&		&	X	&	1.38	&		\\
83	&	5728924	&	18.16	&		&		&	0.06	&	0.06	&	0.09	&	0.5	&	0.82	&	1.26	&	0.27	&		&	X	&	4.32	&		\\
84	&	6065241	&	14.69	&		&		&	0.19	&	1.11	&	3.74	&	43.9	&	49.97	&	38.76	&	-1	&		&	X	&	1.68	&		\\
85*	&	5557932	&	8.14	&		&		&		&		&		&	2321.07	&	1901.52	&	1329.02	&		&	Y	&	UV	&	0.48	&	G Star	\\
88	&	5646749	&	17.01	&		&		&		&	0.13	&	0.26	&	0.58	&	0.7	&	0.39	&	-0.31	&		&	X	&	0.36	&		\\
89	&	6150124	&	7.25	&		&		&		&		&		&	8695.38	&	8431.41	&	6085.98	&		&	Y	&	UV	&	0.6	&	G Star	\\
91	&	5733906	&	11.83	&		&		&	11.55	&	32.96	&	56.18	&	106.39	&	100.9	&	73.71	&	-2.47	&	Y	&	UV	&	0.66	&	Puls. Var$^{16}$	\\
92	&	6550385	&	17.77	&		&		&	274.15	&	661.84	&	74.34	&	0.35	&	0.59	&	0.12	&	0.77	&		&	X	&	1.38	&		\\
93*	&	5568067	&	11.59	&		&		&		&		&		&	2.19	&	3.23	&	3.24	&		&	Y	&	X	&	1.26	&	Cyg A$^{17}$ 	\\
 \enddata
 
 \tablecomments{The nearest KIC counterparts to the X-ray / UV sources, out to a maximum separation of 5\arcsec, and their corresponding \emph{Kepler} magnitudes and fluxes in the FUV and NUV (from \emph{GALEX}), the optical $UBV$ bands (from the \citet{everett12} survey) and the infrared (from 2MASS). All fluxes are shown in mJy. Also given are the $f_X/f_V$ ratios plotted in Figure \ref{fig:optrat}, whether an archived light curve is available in the KIC, whether the X-ray or UV coordinates were used to query the KIC, the angular separation between the query coordinates and the KIC source, and any identifications of the objects in NED or SIMBAD. The references for the IDs are as follows: $^1$\citet{schmidt11}, $^2$\citet{pigulski09}, $^3$\citet{kapanadze13}, $^4$\citet{massaro09}, $^5$\citet{debosscher11}, $^6$\citet{mush11}, $^7$\citet{carini12}, $^8$\citet{gaulme13}, $^9$\citet{scaringi12}, $^{10}$\citet{cannizzo12}, $^{11}$\citet{coughlin11}, $^{12}$\citet{prsa11}, $^{13}$\citet{slawson11}, $^{14}$\citet{catanzaro11}, $^{15}$\citet{uytterhoeven11}, $^{16}$\citet{balona12}, $^{17}$\citet{baade54}.}

 \end{deluxetable}
\clearpage


\begin{deluxetable}{lcccccccc}
 \label{tab:list}
  \tabletypesize{\footnotesize}
 \tablewidth{0pt}
 \tablecaption{Parameters of Spectroscopic AGN\label{t:tab3}}
 \tablehead{
 \colhead{KSw} &
 \colhead{Type} &
 \colhead{\emph{z}} &
 \colhead{Log $M_{BH}$} &
 \colhead{$\lambda_{Edd}$} \\
 \colhead{Num.} &
 \colhead{ } &
 \colhead{ } &
 \colhead{ } &
 \colhead{ }}
 \startdata 

2	&	Sy 1 &	1.506	&	8.53		&	1.474	\\	
3	&	Sy 1	&	1.177	&	8.404	&	0.452	\\
9 (BZB~J1848+4245)	&	BL Lac	&		&		&		\\
20 (Zw~229-15) 	&	Sy 1	&	0.0266	&	7.29	&	0.038	\\
25	&	Sy 1	&	0.609	&	8.898	&	0.042	\\
27	&	Sy 1	&	0.056	&		&		\\
39	&	Sy 1	&	0.533	&	8.382	&	0.113	\\
40	&	Sy 1	&	0.38	&	8.111	&	0.109	\\
55	&	Sy 1	&	0.437	&	8.783	&	0.069	\\
68	&	Sy 1	&	0.36	&	8.553	&	0.026	\\
82	&	BL Lac?	&		&		&		\\
92	&	Sy 1	&	0.182	&	8.473	&	0.017	\\
93 (Cyg A)	&	Radio Galaxy	&	0.056	&	9.398	&	0.003	\\
 
 \enddata
 
 \tablecomments{Spectroscopic type, redshifts, black hole masses and Eddington ratios of spectroscopically-confirmed AGN. Redshifts and black hole masses are not measurable for BL Lacs due to the lack of optical lines. Values of $M_{BH}$ were obtained using the FWHM of the \mgii~$\lambda$2799~\AA~or the H$\beta$ line for high and low redshifts, respectively. Unfortunately, the redshift in source 27 was such that H$\beta$ fell in the dichroic break and was too low for measurement of \mgii. Values of redshift and black hole mass for KSw~93, which is Cygnus~A, were obtained from \citet{tadhunter03}. See Section~\ref{sec:agnspec} for details on how values were obtained.}

 \end{deluxetable}

\newpage

\begin{deluxetable}{lccccccc}
 \label{tab:list}
  \tabletypesize{\footnotesize}
 \tablewidth{0pt}
 \tablecaption{Parameters of Spectroscopic Stars\label{t:tab4}}
 \tablehead{
 \colhead{KSw} &
 \colhead{\emph{$T_{eff}$}} &
 \colhead{Log $g$} &
 \colhead{Spectral} \\
 \colhead{Num.} &
 \colhead{ } &
 \colhead{ } &
 \colhead{Type}}
 \startdata 

1	&	5953	&	4.297	&	F	\\
6	&	4478	&	2.64	&		\\
13	&		&		&	G	\\
14	&	5607	&	4.307	&	G	\\
16	&		&		&	G	\\
19	&	4817	&	4.093	&	K	\\
22	&	4551	&	1.941	&		\\
23	&	8973	&	4.001	&		\\
24	&	4590	&	2.485	&		\\
28	&	5453	&	3.744	&	G	\\
35	&		&		&	F	\\
38	&	4967	&	4.376	&	K	\\
47	&	5899	&	4.232	&	F	\\
49	&	9046	&	4.014	&		\\
52	&	5513	&	3.7	&		\\
53	&		&		&		\\
54	&		&		&	M	\\
57	&	4667	&	3.317	&	K	\\
64	&	3831	&	4.291	&	M	\\
66	&		&		&	M	\\
67	&		&		&		\\
69	&	8742	&	3.929	&	A	\\
71	&		&		&	K	\\
73	&	5570	&	3.906	&		\\
75	&	7292	&	3.566	&		\\
76	&	6165	&	4.282	&	F	\\
78	&		&		&	F	\\
79	&		&		&		\\
81	&	5641	&	3.902	&		\\
84	&		&		&	M	\\
85	&	5617	&	4.21	&	G	\\
89	&		&		&	G	\\
91	&	5241	&	3.688	&	K	\\

 \enddata
 
 \tablecomments{Effective surface temperature, surface gravities, and spectral types indicated by the Palomar spectra for the 33 confirmed stellar KSwAGS sources. The temperatures and gravities are from the \emph{Kepler} Input Catalog (KIC), while the spectral types are given only for those objects with Palomar spectra.}

 \end{deluxetable}

\end{document}